\documentclass[ amsmath,amssymb,notitlepage,]{revtex4-1}

\usepackage{graphicx}
\usepackage{dcolumn}
\usepackage{bm}
\usepackage[caption=false]{subfig}

\usepackage{gensymb}

\begin{document}


\title{Coupled convection and internal gravity waves excited in water around its density maximum at 4$^{\circ}$C}

\author{P. L\'eard}
\altaffiliation{leard@irphe.univ-mrs.fr}
\author{B. Favier}
\author{P. Le Gal}
\author{M. Le Bars}

\affiliation{
 CNRS, Aix Marseille Univ, Centrale Marseille, IRPHE, Marseille, France
}%


\begin{abstract}
Coupled mixed convective and stratified systems are common in natural flows. To study experimentally the associated dynamics, we use a singular property of water: its non-linear equation of state is characterised by a maximum density close to $4^{\circ}$C. By heating the top of a tank at $35^{\circ}$C and cooling the bottom at $0^{\circ}$C, a two-layer configuration spontaneously appears. The convective motion in the bottom layer consists mostly of a large-scale circulation and rising cold plumes. This turbulent flow generates propagating internal gravity waves in the upper stably-stratified layer. PIV measurements are performed and spectral characteristics of the convection and internal gravity waves are presented. An horizontal large-scale reversing flow in the stratified layer is observed which is viscously driven by a third, intermediate layer. This buffer layer is located between the convective and stratified layers and is thermally coupled with the convective one, hence sustaining a strong horizontal shear. Three dimensional direct numerical simulations with geometry and physical parameters close to the experimental ones corroborate our experimental results. 

\end{abstract}

\maketitle

\section{\label{sec:intro}Introduction}
Numerous geophysical and astrophysical flows present a two-layer configuration, with a turbulent convective layer standing above or below a stably stratified one. Examples include planetary atmospheres, stars interiors, and possibly the outermost layer of the Earth liquid core \cite{hirose_composition_2013}. The dynamics of coupled stratified and convective layers are quite complex. Due to the convective motions, internal gravity waves (IGWs) are generated at the interface between the two layers, and propagate in the stratified one. IGWs transport energy and momentum \cite{rogers_internal_2012,bretherton_momentum_1969} from where they are generated to where they are damped. Thanks to their transport properties and non-linear interactions, IGWs are able to generate and sustain large-scale horizontal flows \cite{plumb_interaction_1977,rogers_internal_2012}. Examples of such  large-scale flows driven by IGWs are the oscillations of equatorial zonal winds observed in some planets' atmosphere \cite{fouchet_equatorial_2008,leovy_quasiquadrennial_1991}, including the Earth where it is called the Quasi-Biennial Oscillation (QBO) \cite{baldwin_quasi-biennial_2001}.\\

The IGWs generation by turbulent dynamics has been studied in various experiments. The generation by a single buoyant plume was experimentally studied by Ansong \& Sutherland \cite{ansong_internal_2010}. The penetration of the plume within the stratified layer and the spectral characteristics of the generated IGWs were studied. They found that the peak frequency of the generated IGWs lies in a range close to $0.7N$, where $N$ is the Brunt-V\"ais\"al\"a (or buoyancy) frequency, and that the radial wavenumber is set by the plume cap and not by the width of the plume at the interface.

Deardoff et al. \cite{deardorff_laboratory_1969} and later Michaelian et al. \cite{michaelian_coupling_2002} studied the effect of penetrative convection in a stratified layer in a transient, Rayleigh-B\'enard type experiment. Stratification was initially set up thermally from the top to the bottom of the tank. Then, the fluid was suddenly warmed up at the bottom, triggering Rayleigh-B\'enard convection. IGWs were measured transiently \cite{michaelian_coupling_2002} while the stratified (resp. convective) layer was decreasing (resp. increasing) in size. Eventually, there was no more stratified layer to sustain the propagation of IGWs. 

Townsend \cite{townsend_natural_1964} introduced an original set-up to study the quasi-steady generation of IGWs by Rayleigh-B\'enard convection. Using the fact that water maximum density is around $4^{\circ}$C, a two-layer system is spontaneously generated by cooling the bottom of a tank at $0^{\circ}$C and heating its top above $4^{\circ}$C. The density gradient is unstable at temperature below 4 $^{\circ}$C and stable at temperature above. This creates a self-organising system, with a turbulent convective layer adjacent to a stratified layer. With dye visualisation and temperature measurements, he observed IGWs propagating close to the interface between the two layers. The $4^\circ$C convection was also studied experimentally by Le Gal \cite{legal_penetrative_1997} in a laminar flow, at low Rayleigh number. Convection displayed an hexagonal pattern and viscous entrainment of the fluid above the convective cells was observed. Perrard et al. \cite{perrard_experimental_2013} and Le Bars et al. \cite{bars_experimental_2015} re-investigated this setup in a quasi two-dimensional tank using Particle Image Velocimetry (PIV) and  temperature measurements to obtain detailed data of IGWs generated by the convection. They observed a wide spectrum of waves generated at different frequencies. Favoured frequencies were related to the differential attenuation length of the waves depending on frequency, in good agreement with linear theory. No large-scale flow in the stratified layer was observed in this 2D geometry. Numerical simulations of the same configuration were performed by Lecoanet et al. \cite{lecoanet_numerical_2015}. They demonstrated that IGWs are mainly generated by the Reynolds stresses in the bulk of the convection below the interface, rather than by the displacement of the interface between the two layers. Numerical studies by Couston et al. \cite{couston_dynamics_2017,couston_order_2018,couston_energy_2018} extended these results by considering a generic non-linear equation of state (piecewise linear around an inversion temperature, with adjustable slopes), both in 2D and 3D horizontally periodic geometries. Various flow regimes and the energy transport by IGWs were quantitatively studied. Interestingly, long simulations -- accessible in 2D studies only -- showed, for low Prandtl numbers ($Pr < 1$), a large-scale horizontal flow with reversals in the stratified layer, similar to the QBO phenomenon introduced above \cite{couston_order_2018}. 
\\

Several experiments took interest in the generation and reversal of a large-scale horizontal mean flow in a stratified domain, driven by IGWs. The well-known experiment designed by Plumb and McEwan \cite{plumb_instability_1978}, later reproduced and extended by Semin et al. \cite{semin_generation_2016,semin_nonlinear_2018}, is capable of driving a QBO from the mechanical forcing of a standing wave pattern (\textit{i.e.} two waves with the same frequency and opposite phase speed) in a salty water stratification. With this system, Plumb and McEwan managed to observe few oscillations of the driven large-scale flow before the stratification was destroyed by mixing. The experiment gave results in good agreement with the theory \cite{richard_s._lindzen_theory_1968,lindzen_updated_1972,plumb_interaction_1977}, notably with reversals starting away from the forcing and propagating towards it.
Semin et al. \cite{semin_generation_2016,semin_nonlinear_2018} improved the system by constantly injecting fluid to rebuild the stratification, while removing mixed fluid close to the forcing. This method allowed to run the experiment longer and to study the nature of the bifurcation in the Plumb model which can be either supercritical or subcritical, depending on the dominant dissipative process. In those experimental realisations of the QBO mechanism, the wave forcing remains monochromatic, as opposed to the natural mechanism where it is due to chaotic tropical storms \cite{baldwin_quasi-biennial_2001}. The forcing is driven by interface displacements, as opposed to the observations of \cite{lecoanet_numerical_2015}. Besides, only the stratified layer is modelled. It thus remains a challenge to observe experimentally a large-scale, reversing flow from a turbulent source and a wide range of naturally excited IGWs.
\\

In the present study, we extend the work of Townsend \cite{townsend_natural_1964,perrard_experimental_2013,bars_experimental_2015} in a cylindrical, 3D geometry reminiscent of Plumb and McEwan's set-up \cite{plumb_instability_1978,semin_generation_2016,semin_nonlinear_2018}. Our purpose is threefold: to characterise the generation of IGWs in such a self-organising two-layer system, to quantify the coupling between the layers, and to investigate the possible generation of large-scale horizontal flows. Our experiments are complemented by direct numerical simulations of the same configuration. The experimental setup and numerical model are presented in section \ref{sec:methods}, results are analysed in section \ref{sec:results}, and conclusions and future works are discussed in section \ref{sec:discussion}.

\section{\label{sec:methods}Methods}
\subsection{Experimental set-up}
The set-up consists in a cubic tank whose lateral boundaries are made of 2 cm thick acrylic walls. The bottom boundary is a chrome plated copper plate in which refrigerated fluid is forced to circulate. The top boundary is a commercial, transparent electric heater. The tank inner dimensions are $32 \times32$ cm for the horizontal section and $H=20$ cm in height. Preliminary experiments were conducted in this cubic geometry. Eventually, a cylinder of outer diameter $D = 29$ cm and thickness $e = 0.4 $ cm was added inside the cubic tank, to reproduce the axisymmetric geometry of \cite{plumb_instability_1978,semin_generation_2016,semin_nonlinear_2018}, which seems prone to the development of large-scale flows. We are interested in the flow within the cylinder: the fluid in the gap between the cylinder and the cubic tank follows a similar dynamics and thus imposes to the working fluid a (close to) no-flux lateral boundary condition.

The temperature of the bottom boundary is controlled by water circulating in the copper plate. Water is cooled down by a thermal bath with a fixed temperature set at $-1.25^{\circ}$C. Due to thermal losses in the pipes alimenting the copper plate, bottom tank temperature is $0.2 \pm 0.05^\circ$C. Plate thickness and circulation circuit were optimised so as to ensure a uniform temperature over the whole plate. At the top boundary, the heater is set at a temperature of $35^{\circ}$C. Its temperature control is custom made: a PT100 probe measures the heater temperature in real time, driving through a feed-back loop the input power injected in the heater. This is a very simple and inexpensive system to impose a temperature while having a transparent top boundary, allowing visualisation and velocity measurements by PIV. Nonetheless, it is necessary to point out that the temperature over the heater area is not perfectly homogeneous. Temperature is maximal at the centre where $T \sim 38^{\circ}$C, while the edges are indeed at the requested $T = 35\pm 0.1^\circ$C. This inhomogeneity of the top temperature by $\delta T = 3^{\circ}$C induces slow convective motions below the heater, in a $\sim 2$ cm high layer. By performing an experiment where the whole fluid is stably stratified with an overall temperature gradient similar to the one in the stratified layer studied here, but above the density reversal at $4^{\circ}$C (i.e. bottom boundary set at 10$^{\circ}$C and top boundary at 70$^{\circ}$C), we have checked that those top convective motions have no significant impact on the dynamics of the two-layer system. It is also important to say that despite the thick acrylic wall and the intermediate fluid layer between the cylinder and the tank, the working region is not fully thermally insulated on the sides. Nevertheless, our fully stratified, test experiment has shown no motion within the fluid driven by these lateral losses. 
\\

\begin{figure}[h]

  \includegraphics[scale=.3]{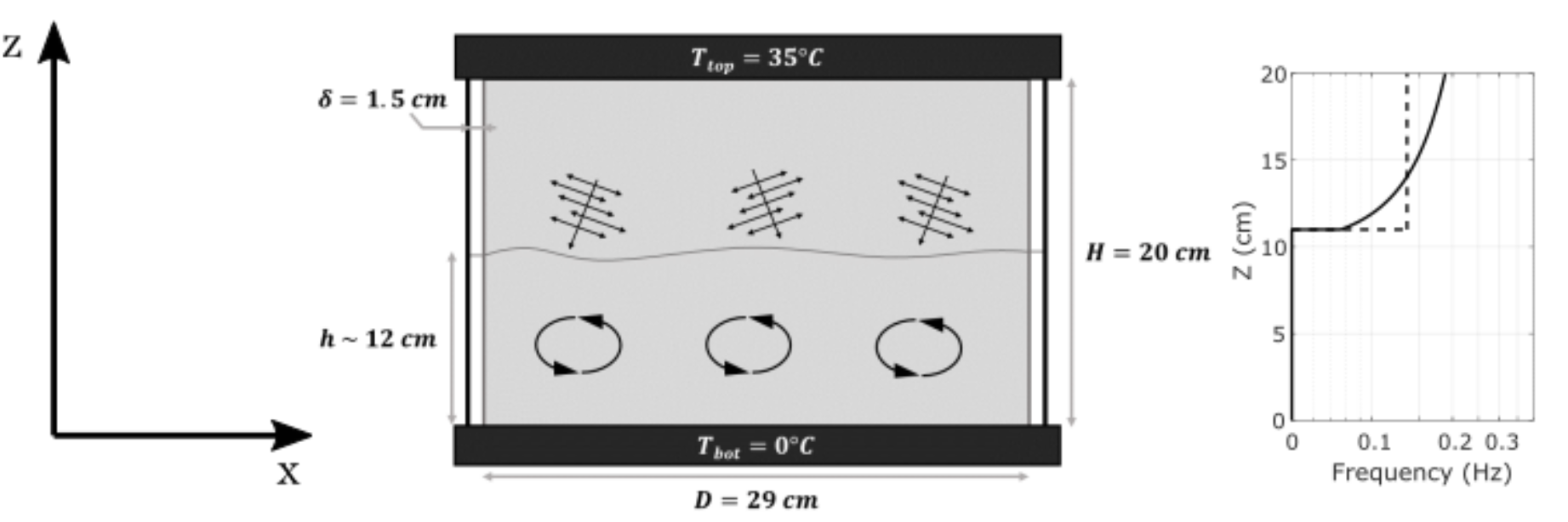}
  \caption{2D sketch of the tank. A cylinder (light grey shaded area) is placed in a larger cubic tank. The system is cooled down at the bottom at 0$^{\circ}$C and heated up at the top at 35$^{\circ}$C. The bottom half is convective with an almost constant density, apart from the bottom boundary layer. The upper half is stably stratified and waves are generated due to the fluid motions in the convective layer. The graph on the right shows the theoretical profile for the buoyancy frequency $N^*$. It is computed considering a linear temperature profile and the equation of state of water (\ref{eq:eos_eau}). The dashed line is the global buoyancy frequency $N$ calculated on the stratified layer. The various length scales are the cylinder diameter $D$, the vertical extent of the tank $H$ and the vertical extent of the convective layer $h$. $\delta$ is the minimal width between the outer square tank and the inner cylinder. The $x$, $y$ and $z$-velocity components are noted $u$, $v$ and $w$ respectively.}
  \label{fig:setup}
\end{figure}

The equation of state of water is non-linear with a  maximum density close to 4$^{\circ}$C (International Equation of State of Seawater, 1980):
\begin{equation}
    \begin{split} 
    \rho(T)= 999.842594+6.793952.10^{-2}T-9.095290.10^{-3}T^2+1.001685.10^{-4}T^3 \\
    -1.120083.10^{-6}T^4+6.536332.10^{-9}T^5.\\
    \end{split}
    \label{eq:eos_eau}
\end{equation}
Thus, due to the imposed temperature at top and bottom boundaries, the bottom part of the tank, between 0$^{\circ}$C and 4$^{\circ}$C, is convectively unstable (see figure \ref{fig:setup}). Cold plumes detach from the bottom plate and rise in the tank due to buoyancy. Reciprocally, ``hot'' fluid sinks from the 4$^{\circ}$C  interface due to gravity. While convective motion takes place in the lower layer, the upper part of the tank, between 4$^{\circ}$C and 35$^{\circ}$C, is stably stratified, with an assumed linear temperature profile at equilibrium. The temperature is indeed linear for an ideal case without thermal losses, assuming that the stratified layer has a bottom boundary at fixed temperature $4^\circ$C and top boundary at $35^\circ$C (\textit{i.e.} constant diffusive flux through the whole layer). However, the density profile is not linear, due to the non-linear equation of state of water. Stratification is characterised by the Brunt-V\"ais\"al\"a frequency $N^* = \frac{1}{2 \pi} \sqrt{-\frac{g}{\rho_0} \frac{\partial \rho}{\partial z}}$. Because of the non-linear equation of state, $N^*$ is not constant with depth, as shown in figure \ref{fig:setup}. For simplicity, we also use below the global buoyancy frequency defined as $N = \frac{1}{2 \pi} \sqrt{-\frac{g}{\rho_0} \frac{\Delta \rho}{\Delta z}}$, where ${\Delta \rho}$ is the global density contrast within the stratified layer of depth ${\Delta z}$.
\\

Before starting the experiment, the bath and the heater are allowed to reach their assigned temperature. Then, the upper half of the tank is filled with water stratified in temperature from 4$^{\circ}$C to 35$^{\circ}$C, using the double bucket technique \cite{oster_density_1965}. The bottom half is filled with water with a temperature close to 4$^{\circ}$C. This filling process is used to avoid tremendously long transient before reaching steady state by thermal diffusion. Typically, we fill the tank at the end of a day and start the measurements the next day in order to let the system reach its equilibrium state over night. Each experiment then lasts four days, with no change in the location of the interface. Note that this steady interface position is the result of the heat budget equilibrium between the convective heat flux extracted from the bottom plate, the diffusive heat flux through the stratified layer, and the lateral heat losses.

To perform PIV measurements, particles are  chosen as small as possible and with a density as close as possible to water density in order to avoid their sedimentation in the stratified layer over the long duration of the experiment. We use fluorescent orange polyethylene micro-spheres whose size ranges from $10 \, \mathrm{\mu m}$ to $20 \, \mathrm{\mu m}$ and density is $1.00 \pm 0.01$~g/cc. The fluorescent property allows us, with a high pass filter on the camera, to remove any laser reflection, significantly enhancing the images quality. The tank is illuminated with a green laser $532$~nm. Power is set at $1$~W. 
We perform side view PIV to measure convection and IGWs spectral characteristics, and top view PIV to observe the large scale flow and its fluctuations over time. The camera used for the side view PIV is a HiSense Zyla $2560 \times 2160$ pixels recorded on 12 bits. Acquisition rate is $2$~Hz with $100$~ms exposure time. Typical acquisition time for spectral characteristics is $50$~min.
For the top view, we use a Point grey camera $1920\times1080$ pixels on 8~bits. Exposure time is $1$~s, acquisition rate $0.1$~Hz and acquisition time $8$~hours. Captured movies are processed either by the DantecDynamics software DynamicStudio for the side view or by DPIVSoft \cite{meunier2003analysis} for the top view. Both are resolved into $32\times32$ pixels boxes with 50\% overlapping. 

\begin{figure*}[h]
    \centering
    \includegraphics[scale = .45]{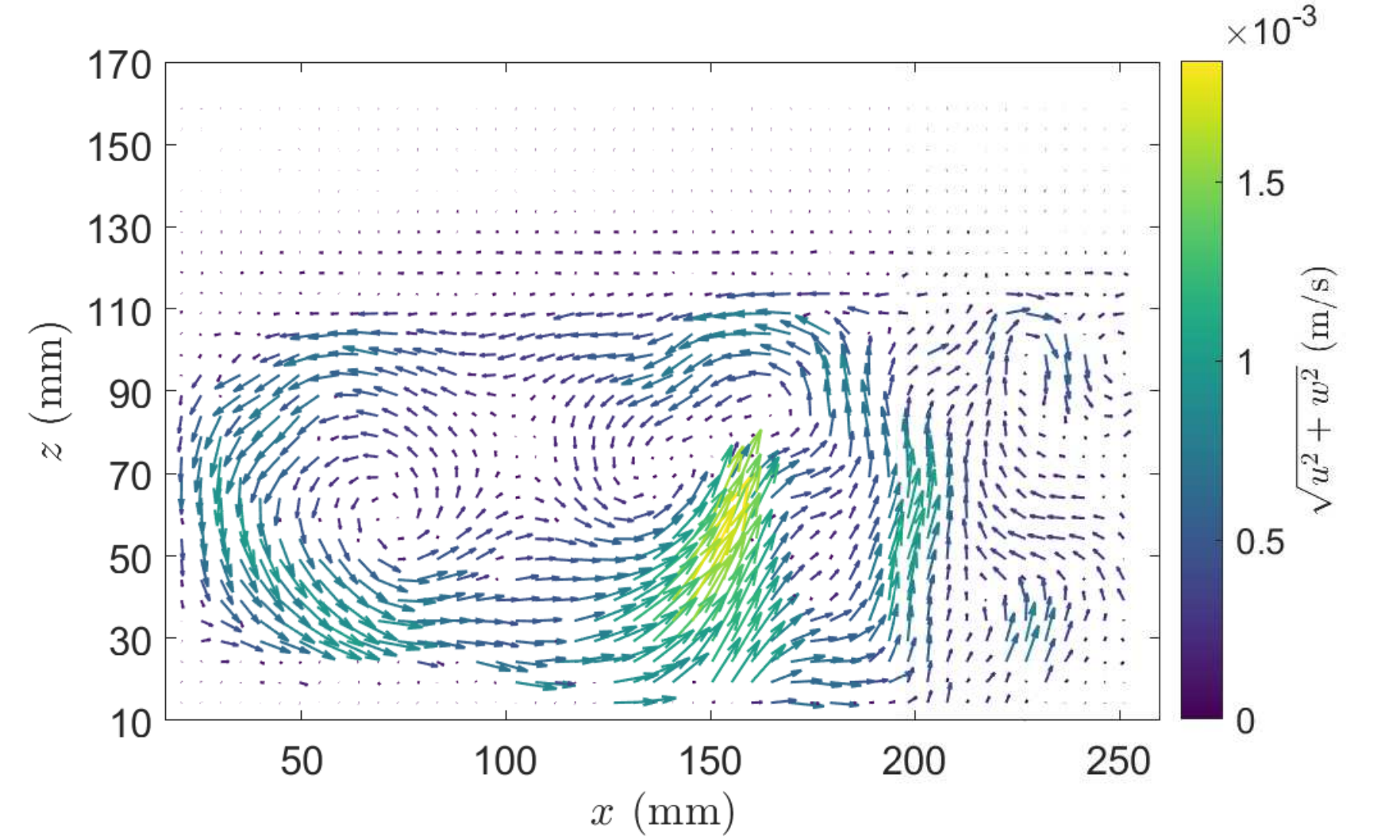}
    \hspace{.1cm}
    \includegraphics[scale = .45]{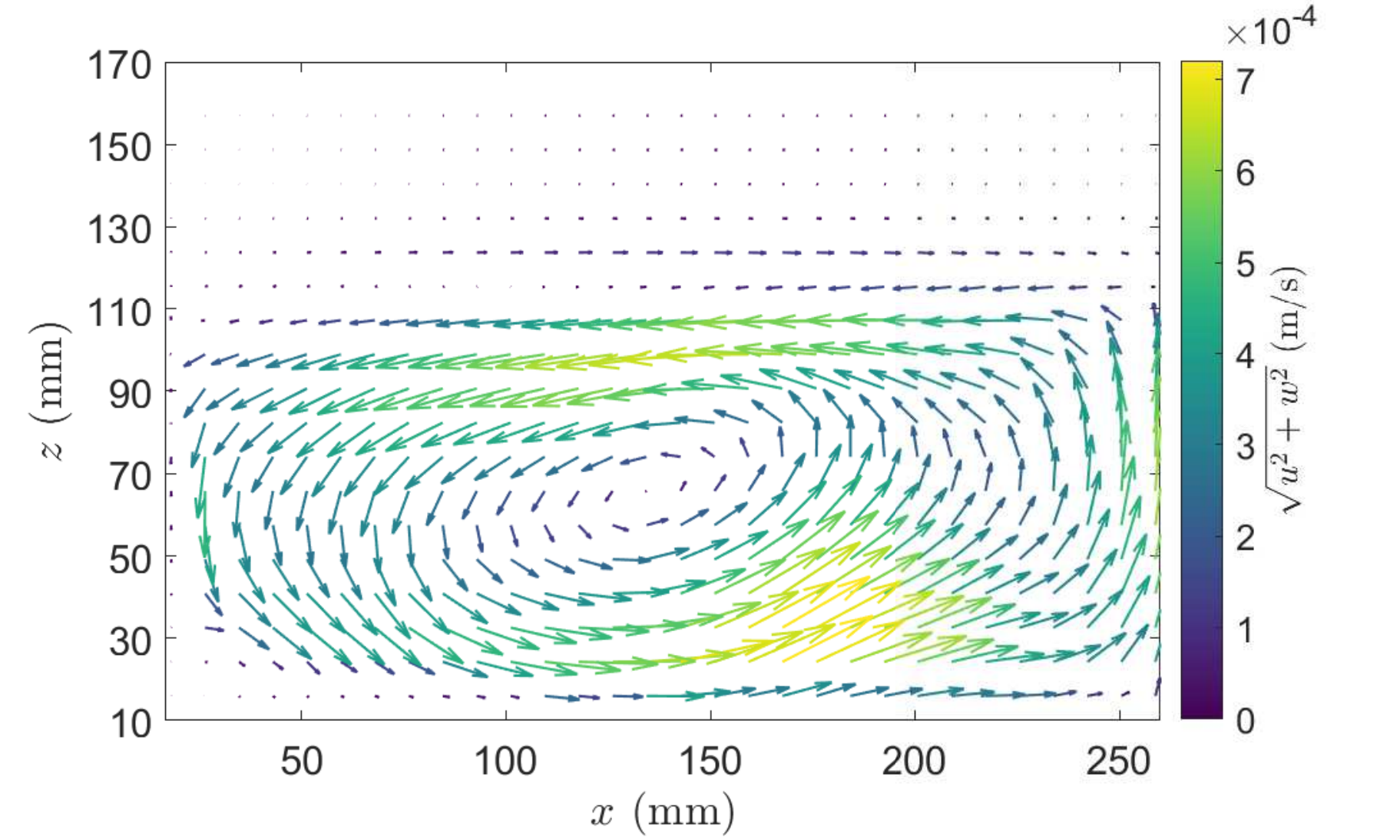}
    \caption{(Left) Instantaneous velocity field. An ascending plume is visible at $x=150$ mm, and transported by the large-scale circulation. (Right) Large-scale circulation in the convective layer obtained by time-averaging velocities over a 50 minutes signal. The large-scale circulation is a counter clockwise cell. No inversion of the circulation has been seen in our experiment. The velocities under $z=10$ mm are noisy and thus not shown here. Maximum instantaneous velocities are $2.5$ times bigger than the maximum averaged velocities. The left edge of the cylinder is located at $x=19$~mm and the centre of the cylinder is at $x=160$~mm. Approximately $40$~mm of the right side of the cylinder is not captured with our PIV measurments.}
    \label{fig:LSCexpe}
\end{figure*}

Side view PIVs are performed in the middle of the tank at $y=16$~cm in a laser sheet crossing the cylinder along its diameter. This is the case for all figures shown in the $(x,z)$ plane and thus not mentioned in the results section. The vertical fields (see an example in figure \ref{fig:LSCexpe}) do not show the whole $(x,z)$ plane (where the origin $(0,0)$ is located in the bottom left corner of the cubic tank): it was indeed chosen to zoom in, in order to have the best resolution for the very weak motions in the stratified layer. The interface between the layers is localised between $11 \mathrm{~cm} \leqslant z \leqslant 12$~cm. Typical Rayleigh number for the convection based on this depth is $Ra = 7 \times 10^6$, and the global Brunt-V\"ais\"al\"a frequency is $N = 1.35 \times 10^{-1}$~Hz.

To observe the large-scale flow, horizontal views at different heights are performed. A linear translation axis platform from Igus, driven by homemade software, is used to automate the process. With two mirrors, one fixed at the exit of the laser head and the other fixed on the translating platform, it is possible to sweep along one direction with the laser sheet. We typically make measurements during 15~s in each of 11 successive planes separated by 0.5~cm. The full scan duration is about 3~min, and is repeated during at least 8~hours.
\\

The cylindrical geometry described here differs from the cylindrical shell geometry in \cite{plumb_instability_1978,semin_generation_2016, semin_nonlinear_2018}. We first tried to work in that annular geometry by adding a second, smaller cylinder in our cubic tank. Three different gap sizes were tested but none showed any interesting dynamics. Indeed, the convection was too confined in the shell to provide an efficient chaotic excitation, and IGWs did not propagate well within the stratification, attenuated quite fast by wall friction. During these tests, we observed the most interesting dynamics within the innermost cylinder, so we decided to use that geometry. The point of the cylindrical shell geometry is that it is a close analogue to the equatorial band of the stratosphere where QBO takes place. By working in a cylinder, the geometry analogy is lost but the physics of the problem remains the same, still a priori allowing for large-scale, reversing, axisymmetric horizontal flows.

\clearpage
\subsection{Numerical model}\label{sec:methods_num}

To complement the experiments, we also performed Direct Numerical Simulations (DNS) of the same configuration. We solve the Navier Stokes equations using a non-Oberbeck Boussinesq model. The density variations are consider small compared to the reference density $\frac{\Delta \rho }{\rho_o} \ll 1$. Therefore, density fluctuations only appear in the buoyancy force. However, temperature variations affect the value of the thermal expansion coefficient to account for the non-linear equation of state of water. Variations of the thermal diffusivity $\kappa$ and kinematic viscosity $\nu$ are neglected. Governing equations are given in dimensionless form by:
\begin{align}
\label{eq:momentum}
\frac{\partial\bm{u}}{\partial t}+\bm{u}\cdot\nabla\bm{u} = & -\nabla P+Pr\nabla^2\bm{u}-Pr Ra \ \theta^2 \bm{e}_z \\
\label{eq:heat}
\frac{\partial \theta}{\partial t}+\bm{u}\cdot\nabla\theta = & \ \nabla^2 \theta \\
\label{eq:mass}
\nabla\cdot\bm{u} = & \ 0
\end{align}
where we used the depth $H$ of the container as a unit of length, the thermal diffusive timescale $H^2/\kappa$ as a unit of time and the difference between the bottom and the inversion temperature $T_0 - T_i$ as a temperature scale. These equations are characterised by the Prandtl number $Pr=\nu/\kappa$, the global Rayleigh number $Ra=\alpha g (T_0-T_i)^2 H^3/(\nu\kappa)$ and the imposed dimensionless top temperature $\theta_{top}=(T_{top}-T_i)/(T_0-T_i)$.
Note that the quadratic temperature term in the momentum equation is a direct consequence of the nonlinear equation of state of water given by equation~\eqref{eq:eos_eau}, which is approximed in our model by the quadratic equation $\rho(T) \approx \rho_0 (1-\alpha(T-T_i)^2)$.
The coefficient $\alpha$ is not the usual thermal expansion coefficient but has a unit of $(\degree \mathrm{C})^{-2}$ and is given by $\alpha\approx8.1\times10^{-6}(\degree \mathrm{C})^{-2}$ (see also \cite{lecoanet_numerical_2015}).

We consider a cylindrical fluid cavity of diameter $D=3H/2$ as in the experiments.
Both horizontal plates are assumed to be no-slip and with fixed temperature.
The side wall is assumed to be no-slip and perfectly insulating.
This is of course not the case in the experiment, for which lateral heat losses are inevitable and top temperature is not exactly constant, but the objective is to check whether the conclusions drawn from the experimental results are robust and do not depend on these effects.
Since the experiment runs with water, we use $Pr=7$.
The Rayleigh number of the experiment is $Ra=7 \times 10^7$ while its dimensionless top temperature is $\theta_{top}=-7.75$. If we were to run the simulation with these parameters, the interface will be located very close to the top boundary. It is not the case in the experiment because of the lateral heat losses, which tend to reduce the effective Rayleigh number. For that reason, and instead of taking into account these losses as in \cite{lecoanet_numerical_2015}, we kept the insulating lateral boundaries and use the slightly adjusted parameters $Ra=10^7$ and $\theta_{top} = -11$ instead, which leads to an interface located approximately at $z\approx120$~mm, as in the experiment. The Rayleigh number could not be lowered under $10^7$ in order to keep the convective flow turbulent enough, thus we had to increase the top temperature to have the interface located at $z\approx120$~mm.

We perform DNS of equations~\eqref{eq:momentum}-\eqref{eq:mass} using the spectral element solver Nek5000 \citep{Nek5000}.
The global geometry is decomposed into hexahedral elements, with vertical refinement close to the horizontal boundaries and around the mid-plane where the inversion isotherm is expected to be located.
Velocity, buoyancy and pressure variables are represented as tensor product Lagrange polynomials of order $N$ and $N-2$ based on Gauss or Gauss-Lobatto quadrature points.
The total number of grid points is given by $\mathcal{E}N^3$ where $\mathcal{E}$ is the number
of elements.
For all the results discussed in this paper, the number of elements is $\mathcal{E}=8960$ and we use a polynomial order of $N=11$. Numerical convergence was checked by increasing the polynomial order $N$.
Time integration is performed with a third-order mixed implicit-explicit scheme.
The simulations are initialised with a small amplitude buoyancy perturbation and a temperature profile varying linearly between the top and bottom boundaries.
Simulations are run until a quasi-steady state is reached, after which data is accumulated to compute statistics.

\section{\label{sec:results}Results}
\subsection{\label{sec:results_exp}Experiments}
\subsubsection{\label{sec:results_conv}Convection}
\begin{figure}[h]
    \centering
    {\label{fig:panacheup}\includegraphics[scale=0.43]{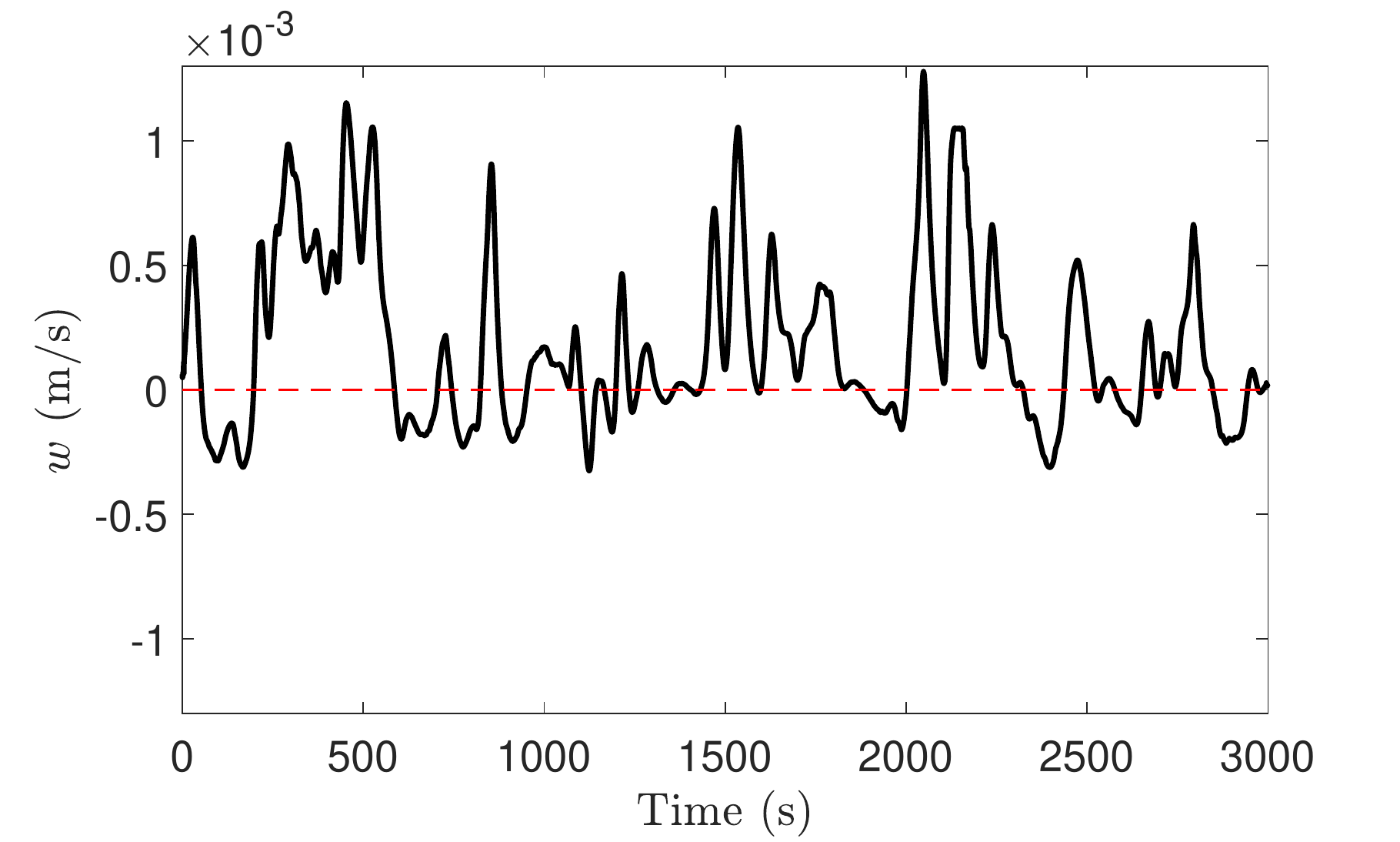}}
    \hspace{.2pt}
    {\label{fig:panachedown}\includegraphics[scale=0.43]{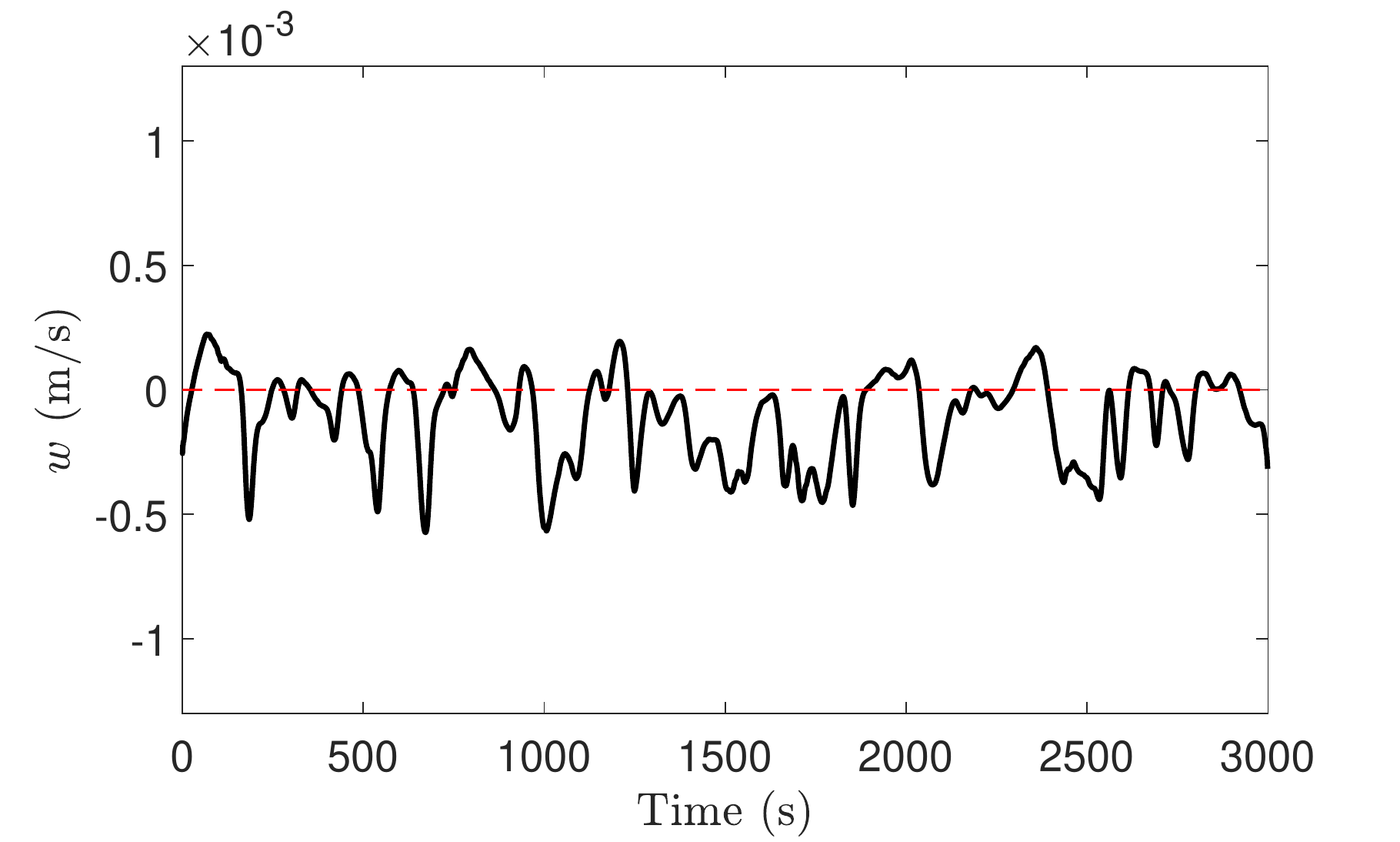}}
    \caption{Time evolution of the vertical velocity $w$ within: (Left) upward plumes at $x = 200$ mm, $z = 45$ mm, (Right) downward structures at $x= 100$ mm, $z = 95$ mm. }
    \label{fig:panache}
\end{figure}

PIV side view is used to quantify horizontal and vertical velocities in the convection zone. Examples of vertical velocities measured at one point in a given location are shown in figure \ref{fig:panache}, for both ascending cold and descending hot structures. Measurements are consistent with the numerical simulations \cite{lecoanet_numerical_2015,couston_dynamics_2017} showing intense, localised, cold rising plumes and more diffusive descending plumes. Moreover, these structures are advected by a large-scale circulation encompassing the whole convective layer, as shown in figure \ref{fig:LSCexpe}.

\begin{figure*}[b]
    \centering
    \includegraphics[scale = .43]{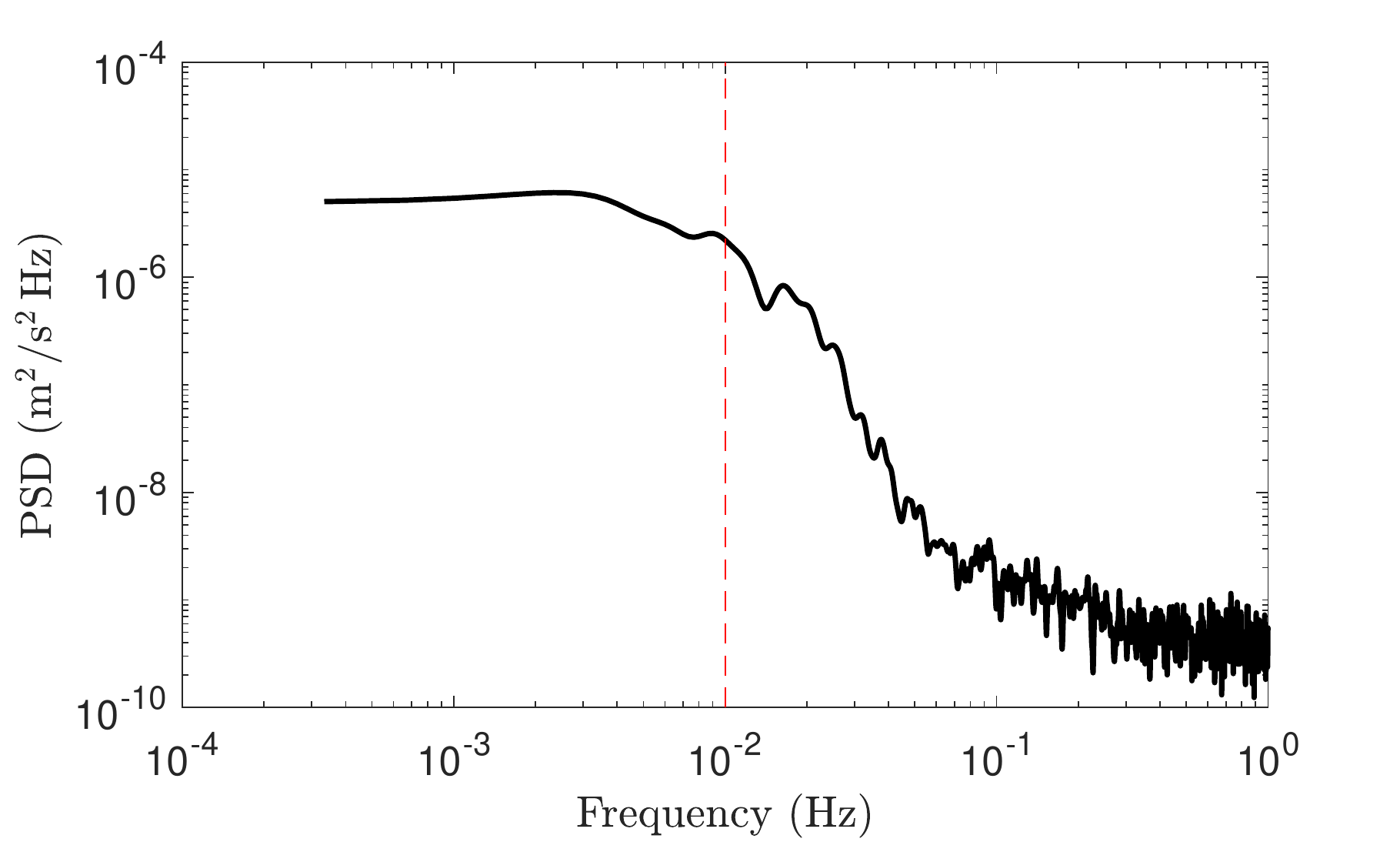}
    \hspace{.1cm}
    \includegraphics[scale = .43]{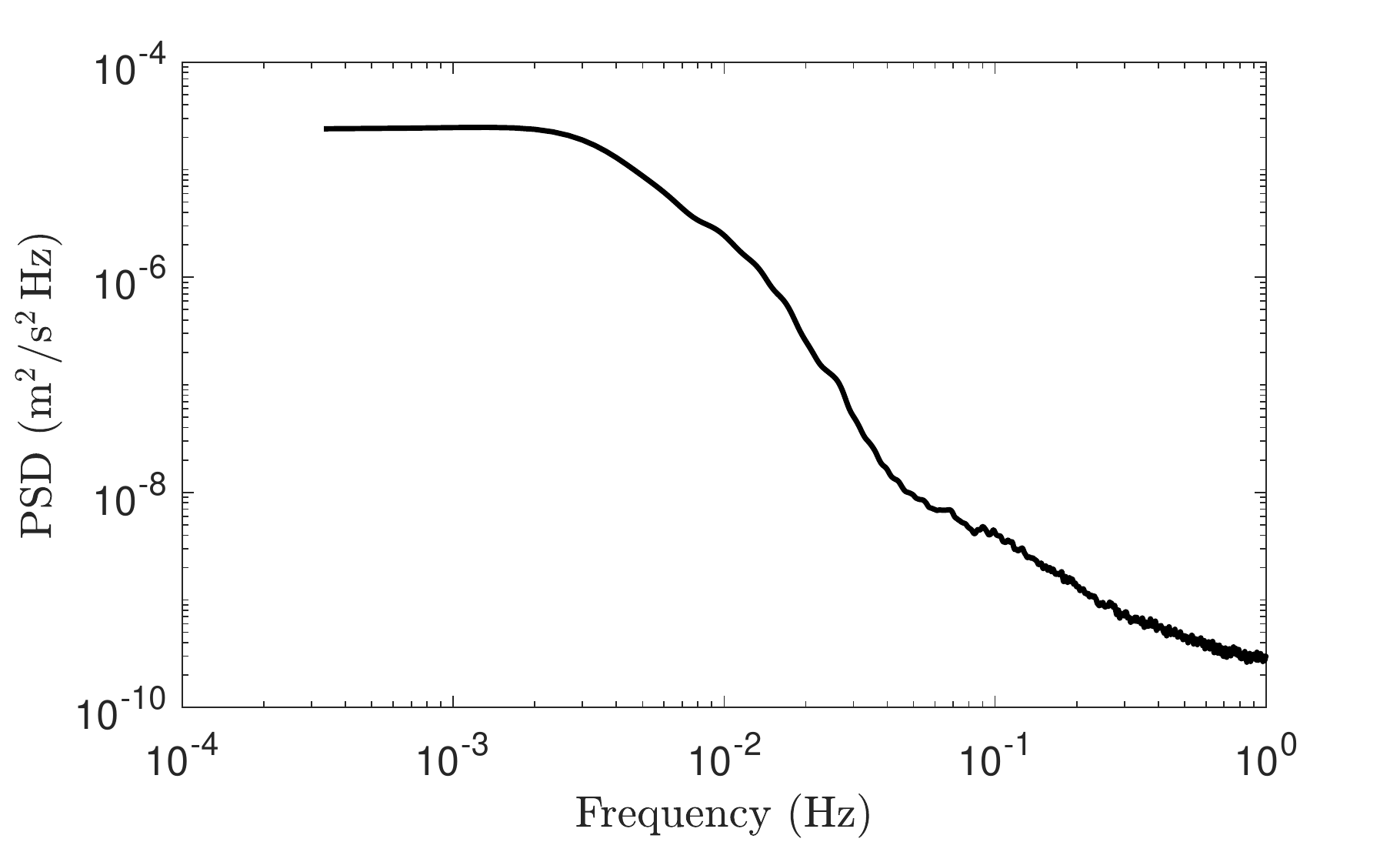}
    \caption{PSD for the vertical velocity fluctuations. (Left): PSD computed at a single point $x=100$~mm, $z=95$~mm (signal shown in figure \ref{fig:panache} right). The plume forcing frequency can be seen around $f = 10^{-2}$~Hz (red dashed line). (Right): PSD spatially averaged over the whole convective cell in the measured $(x,z)$ plane (all points above $z=10$~mm and below $z=110$~mm). }
    \label{fig:spectreconv}
\end{figure*}

Spectral analysis is performed to extract power spectral density (PSD) from the velocity signals. Figure \ref{fig:spectreconv} shows the PSD of the convection vertical velocity $w$. For the two panels, the spectrum is flat with a lot of energy for low frequencies, then the energy drops above some cut-off frequency. Left panel of figure \ref{fig:spectreconv} shows the vertical velocity PSD at a single point close to the top of the convective layer. A small peak can be seen close to $f = 10^{-2}$~Hz. This is the quasi-periodic signal of the plumes dropping from the top thermal boundary layer. The theoretical characteristic time of convection can be computed from \cite{gortler_convection_1966}:
\begin{equation}
    \tau = \frac{h^2}{\pi \kappa}\left(\frac{\Delta T}{\Delta T_{local}} \times \frac{Ra_c}{Ra}\right)^{2/3}
\end{equation}

with $h$ the height of the convective layer, $\kappa$ the thermal diffusivity, $\Delta T$ the temperature difference between the top and bottom of the convective domain, $\Delta T_{local}$ the temperature difference between the top and bottom of the thermal boundary layer, and $Ra_c$ the critical Rayleigh number. The critical Rayleigh number in the presence of free and solid interfaces and for fixed temperature is $Ra_c = 1100.65$. For our experiment, the characteristic time is $\tau = 96$~s, thus characteristic frequency is $1/ \tau \sim 10^{-2}$~Hz, which is close to the observed peak in the left panel of figure \ref{fig:spectreconv}. At frequencies lower than this characteristic frequency, the spectrum is flat. This is explained by the combined effect of the randomness of the plumes (see figure \ref{fig:panache}) and of the slow fluctuations of the large-scale circulation. Right panel of figure \ref{fig:spectreconv} shows the PSD of vertical velocities, averaged over the whole convective cell in the $(x,z)$ plane. It shows a similar trend, with a lower cut-off frequency compared to right panel spectrum. Actually, the plumes signal is more localised and less intense on average than the large-scale circulation signal, which hence dominates the space-averaged PSD.

\begin{figure}
    \centering
    \includegraphics[scale = .6]{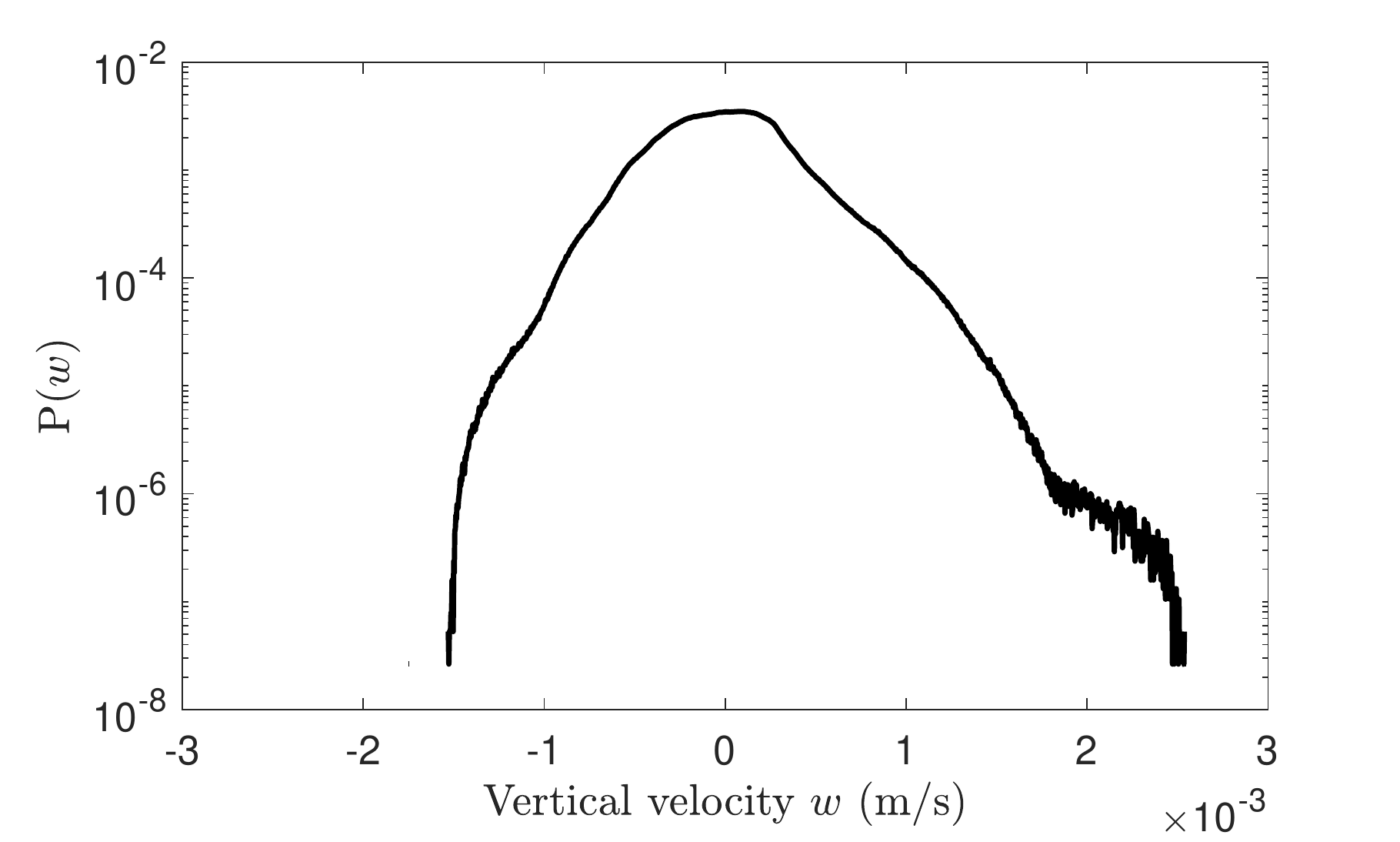}
    \caption{Probability density function of the vertical velocities in the convective layer. All PIV points under $z=110$~mm have been used to compute the PDF.}
    \label{fig:pdf}
\end{figure}

The probability density function (PDF) of the vertical velocities in the whole convective layer $\mathrm{P}(w)$ is computed and shown in figure \ref{fig:pdf}. It is normalised such that $\int \mathrm{P}(w)\mathrm{d}w = 1$. The PDF describes important features of the convection: it is skewed towards positive values, with positive velocities reaching higher magnitude than negative velocities, \textit{i.e.} the ascending plumes are stronger than the descending structures. However, the central part of the PDF is close to gaussian profile. The distribution obtained here is in good agreement with the probability density function computed in an idealised 2D numerical model by Couston et al. \cite{couston_dynamics_2017}. Note that this asymmetry is specific to our model, for which the usual upside-down symmetry in Boussinesq Rayleigh-B\'enard convection is broken.

\subsubsection{\label{sec:results:buffer}Buffer layer}
An intermediate layer (we name it the buffer layer in the following) is present between the convective layer and the stratified layer. It was first reported in the quasi-2D 4$^{\circ}$C convection experiment by Perrard et al.  \cite{perrard_experimental_2013}. Their temperature measurements showed that the buffer layer corresponds to the area where the temperature goes from 4$^{\circ}$C to 8$^{\circ}$C. This actually corresponds to the overshooting region for rising cold plumes (note that this type of convection is called "penetrative convection" because of this effect). Indeed, since the density of water is close to quadratic around $4^{\circ}$C, densities at e.g. $0^{\circ}$C and $8^{\circ}$C are the same, and the 8$^{\circ}$C isotherm is the theoretical maximum height reachable by an ascending cold plume at $0^{\circ}$C in the absence of thermal diffusion. Simultaneously, the overall density profile between $4^{\circ}$C and $8^{\circ}$C is stable, as in the stratified layer above $8^{\circ}$C. The buffer layer is thus a very specific location supporting simultaneously convective overshooting motions and IGWs, as observed with PIV measurements \cite{perrard_experimental_2013}. 

\begin{figure}[h]
    \centering
    \includegraphics[scale = .6]{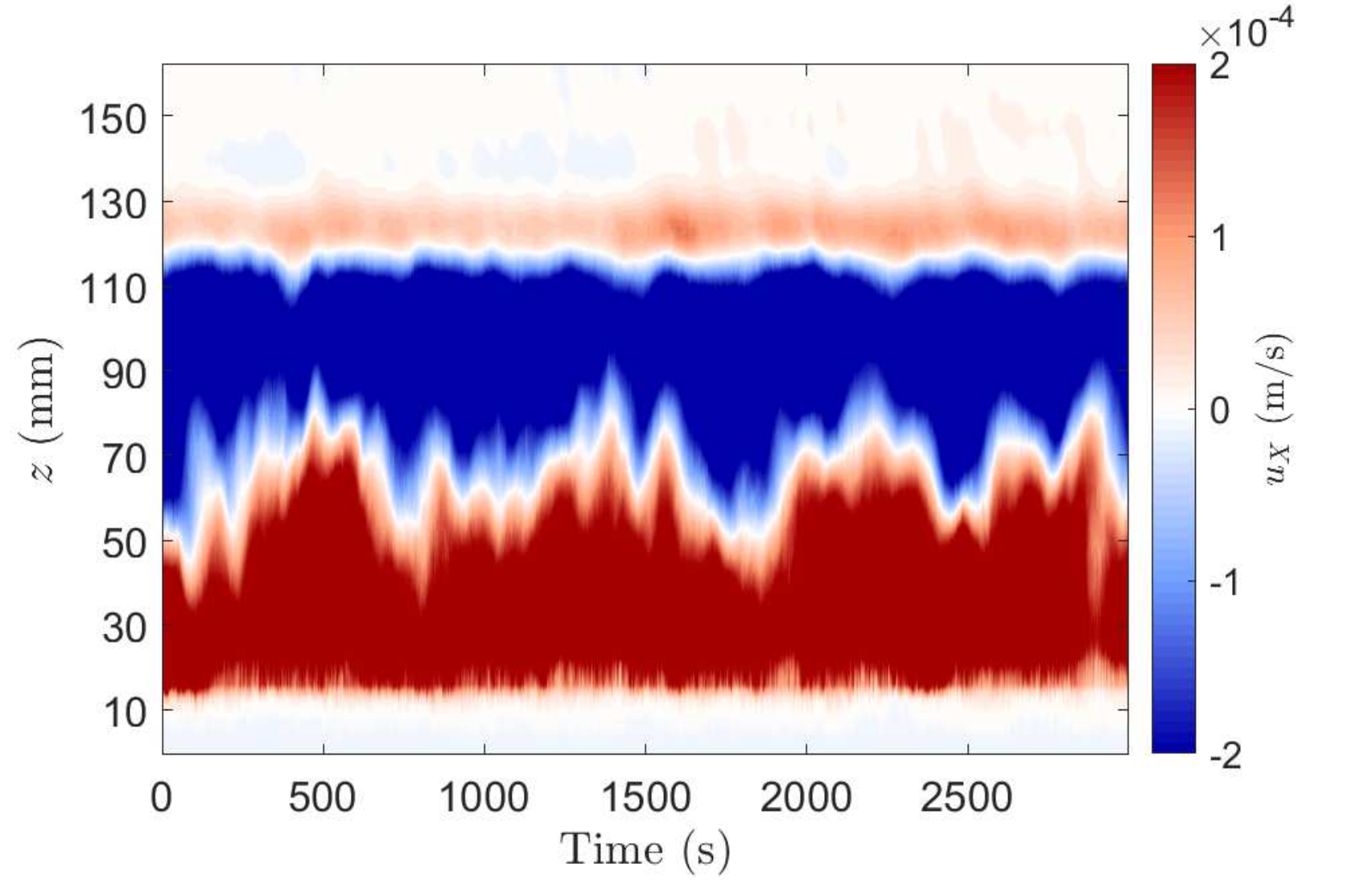}
    \caption{Time evolution of the horizontal average of the horizontal velocity, noted $u_X$. Red (resp. blue) regions correspond to mean flow going towards the right (resp. left).}
    \label{fig:bufferlayer}
\end{figure}

\begin{figure}[htb]
    \centering
    \includegraphics[scale = .75]{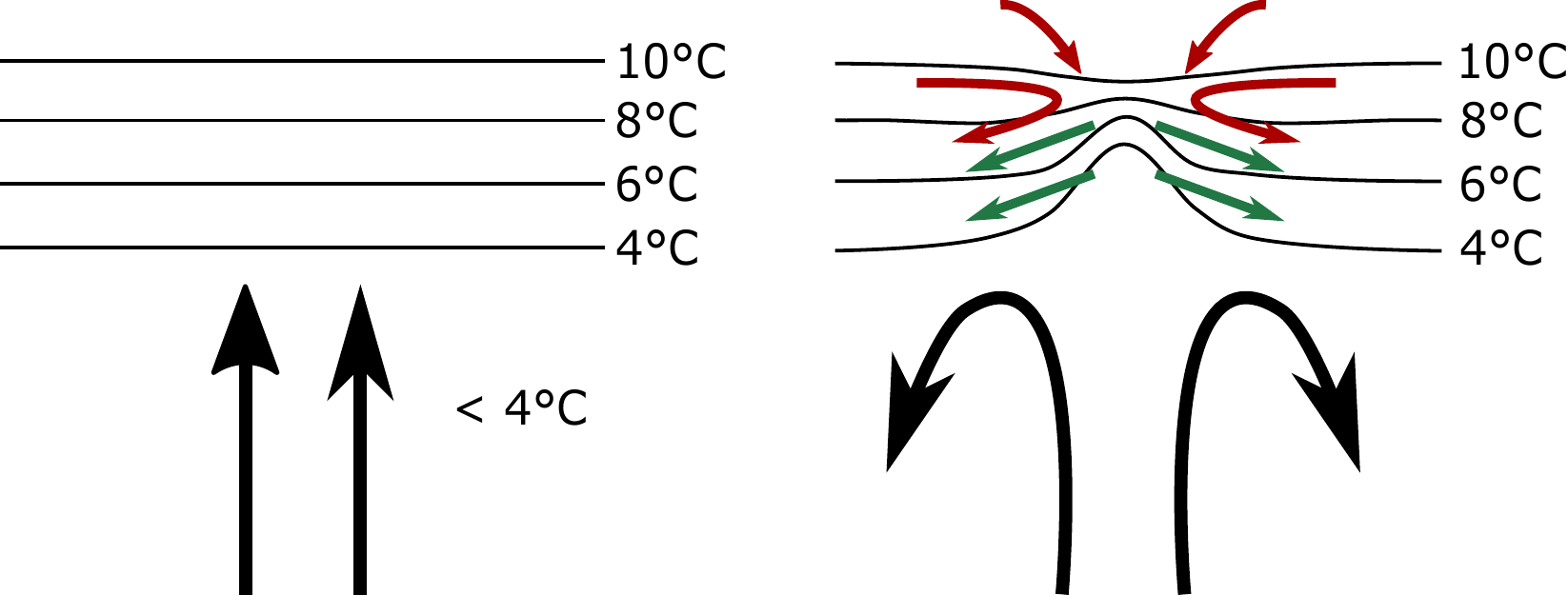}
    \caption{Sketch of the thermal coupling between the convective and buffer layers. On the left, a cold plume moves upwards towards the interface between the two layers. On the right, isotherms are deflected due to the impact.}
    \label{fig:schema_couplage}
\end{figure}

To complete the description of this buffer layer now using velocity measurements, we plot in figure \ref{fig:bufferlayer} the spatio-temporal graph of the horizontal average of the horizontal velocity $u$. The graph exhibits a strong shear around $z=120$ mm. Since the fluid is going in opposite directions above and below $z=120$ mm with a sharp interface, viscous entrainment by the convective layer is excluded. A special kind of thermal coupling might explain the observed dynamics, as sketched in figure \ref{fig:schema_couplage}. Indeed, when a cold ascending plume from the convection zone reaches the interface and overshoots in the buffer region, its associated velocity perturbation dissipates more rapidly than its temperature perturbation. Due to gravity, the distorted part of the buffer region wants to sink back to its initial state (pictured by the green arrows), while the fluid above the buffer layer moves towards the impact point of the plume to take the place of the falling water (pictured by the red arrows). The buffer layer then needs some compensating fluid from the convective layer. This mechanism works when the velocity perturbation of the plume at the interface dissipates more rapidly than the thermal perturbation, hence for a Prandtl number $Pr \geq 1$. One might expect the shearing zone to decrease in size and amplitude when thermal diffusion increases (i.e. when the Prandtl number decreases), since the overshooting rising cold plume will then equilibrate thermally during its ascent more rapidly. This may explain why no interfacial shear was reported in the systematic numerical study of \cite{couston_dynamics_2017,couston_order_2018} where $Pr \leq 1$. Global temperature field measurements (using e.g. Temperature Dependent, Laser Induced Fluorescence) are now required to confirm or infirm the proposed model, but those are beyond the scope of the present paper. Note that by extension, we call "buffer layer" in the following the region including the $T=4^\circ$C to $T=8^\circ$C overshooting region and the shear region. In the experiment, the shear region extends from $z=120$~mm to $z \approx135$~mm.

\clearpage
\subsubsection{\label{sec:results_waves}Internal gravity waves}
\begin{figure}[h]
    \centering
    \includegraphics[scale = .83]{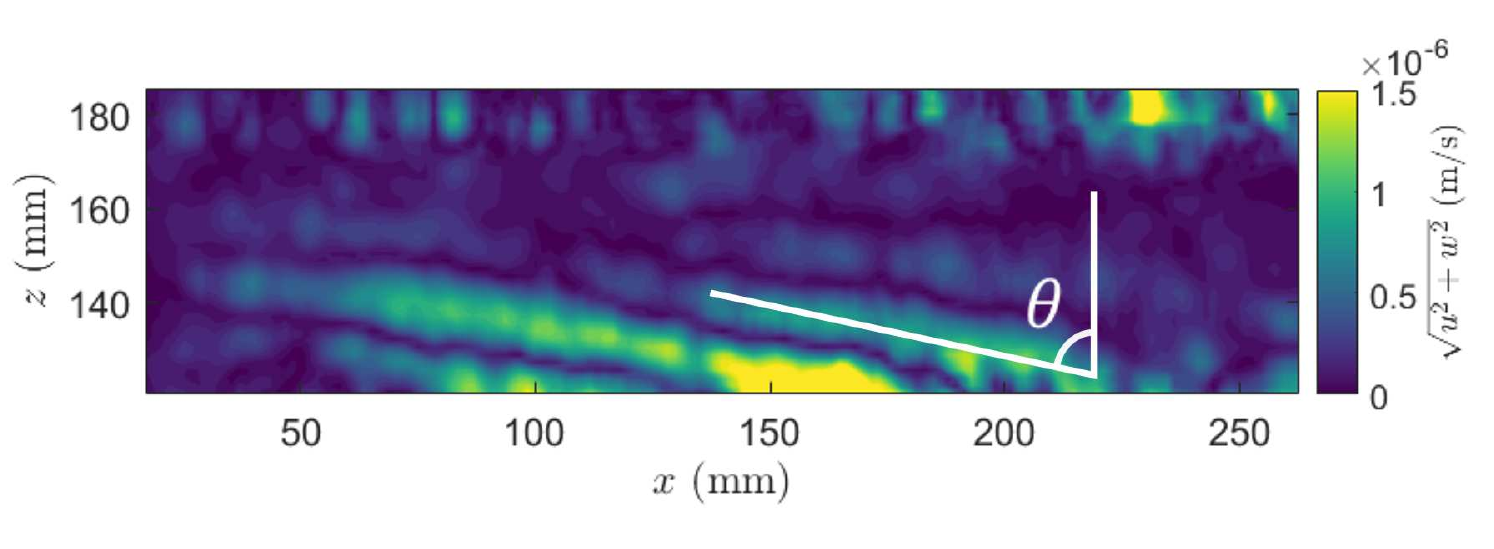}
    \hspace{.1cm}
    \includegraphics[scale = .53]{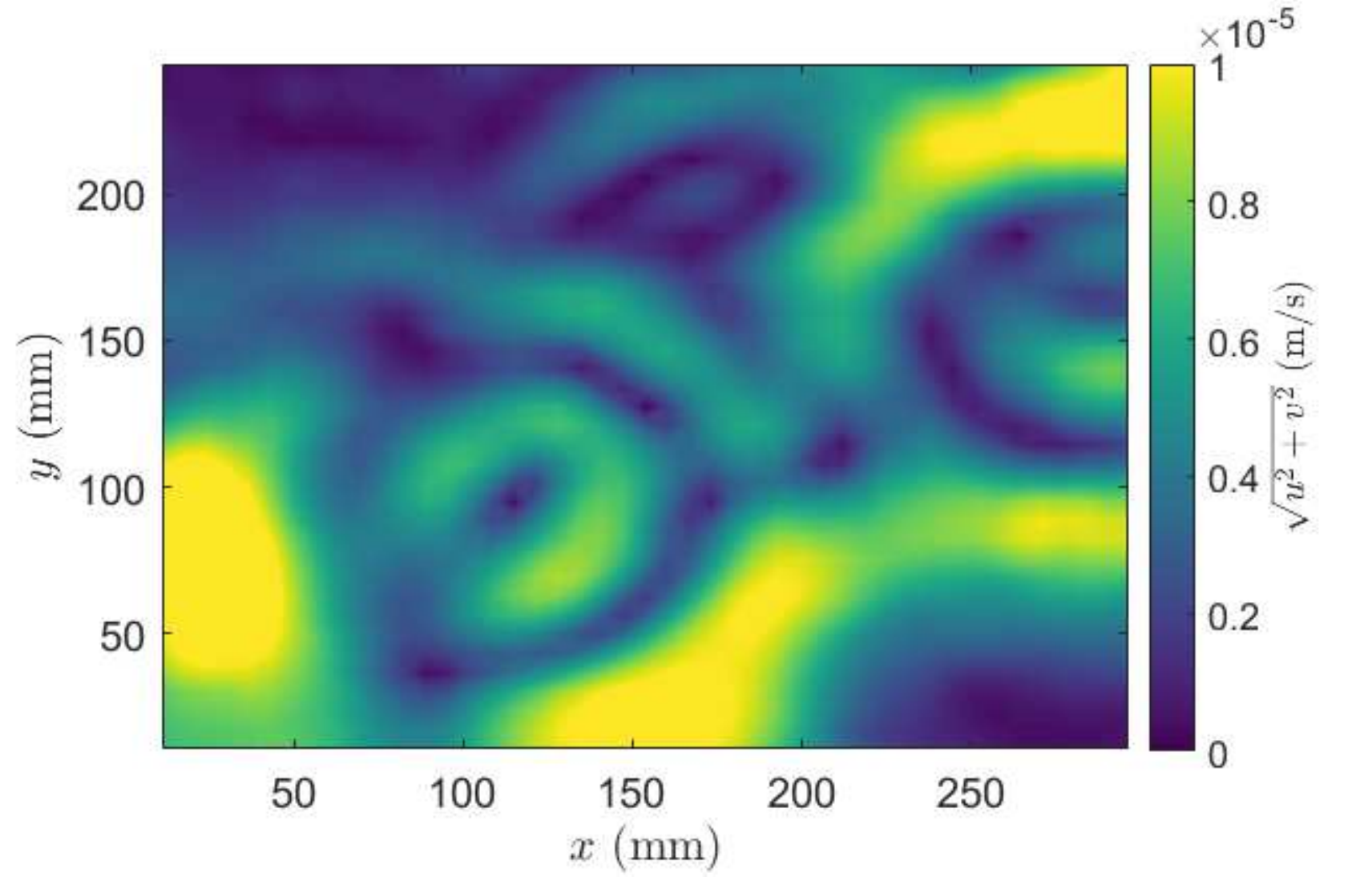}
    \caption{Velocity fields showing IGWs propagating. (Top) Velocities in $(x,z)$ plane. The signal is frequency-filtered to enhance the visualisation of oscillatory motions: only frequencies between $0.02$ and $0.03$ Hz are shown, propagating at an angle of roughly $75^\circ$ with the vertical. The angle of propagation is the angle between the constant phase line and the vertical. (Bottom) Velocities in $(x,y)$ plane located at $z \approx 125$~mm. In the $(x,y)$ plane, IGWs take the form of oscillating rings. Note that this figure is from a previous experiment without any internal cylinder and is therefore only displayed here as an illustration of the IGWs seen from above.}
    \label{fig:ondeN}
\end{figure}

 The convective motions induce a Reynolds stress at and below the interface which generates IGWs propagating in the stratified area \cite{lecoanet_numerical_2015}. An example is shown in figure \ref{fig:ondeN}. The vector field has been frequency filtered in the band $[0.02-0.03]$ Hz to isolate a single propagating wave train. We can measure an angle close to $\theta \simeq 75^{\circ}$ between contant phase lines and the vertical. This observation is in good agreement with the inviscid dispersion relation $\omega = \pm N \mathrm{cos}(\theta)$, which relates the frequency and the propagation angle of IGWs. Indeed, at $z=120$~mm, $N \sim 0.1$~Hz, thus $\theta = \mathrm{cos^{-1}}\left(\frac{\omega}{N}\right) = 78.5^\circ$. The motion within the stratified area is a superimposition of many such IGWs oscillating at different frequencies.

\begin{figure*}[t]
    \centering
    \includegraphics[scale=0.55]{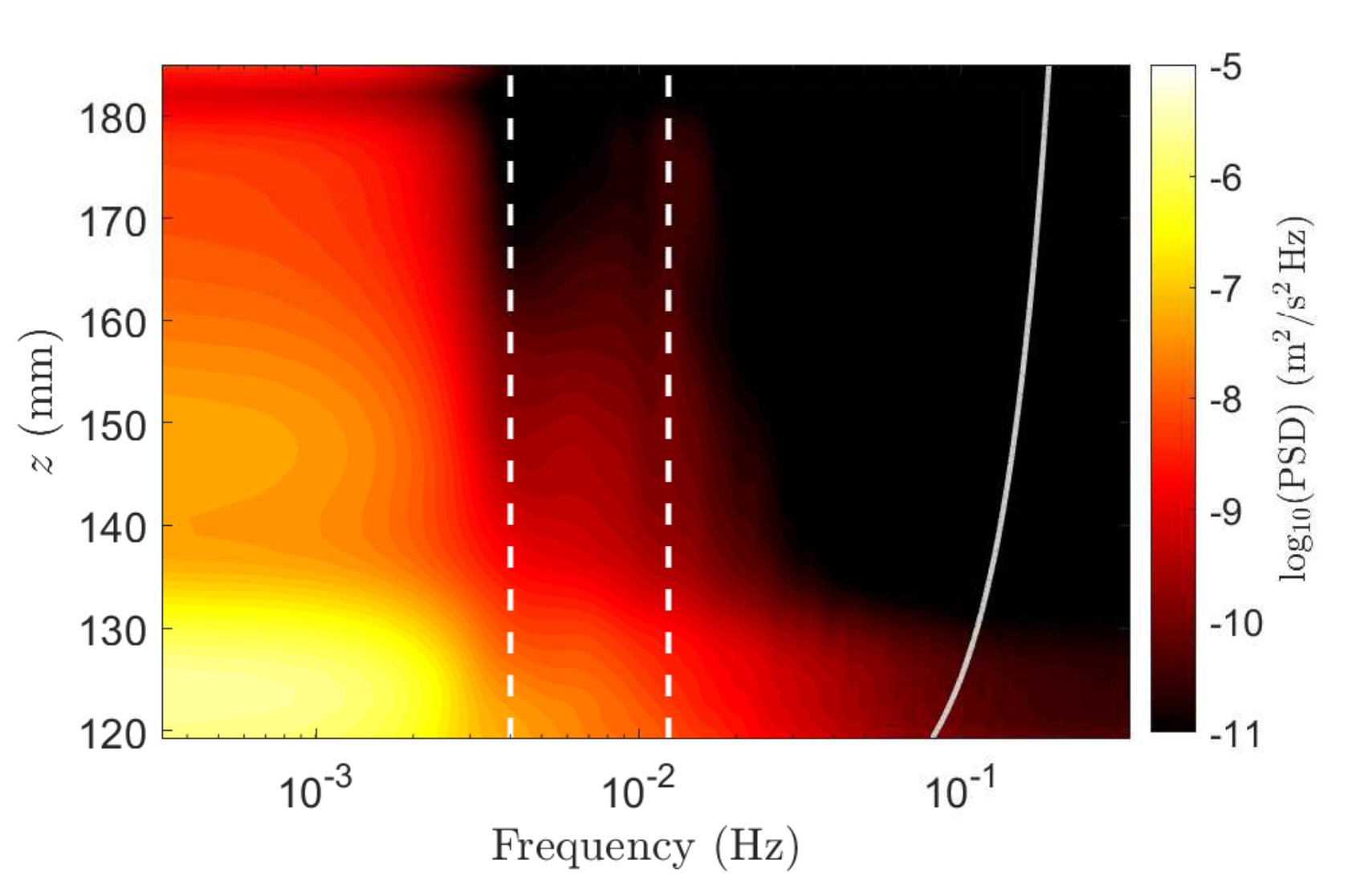}
    \hspace{.1cm}
    \includegraphics[scale =.55]{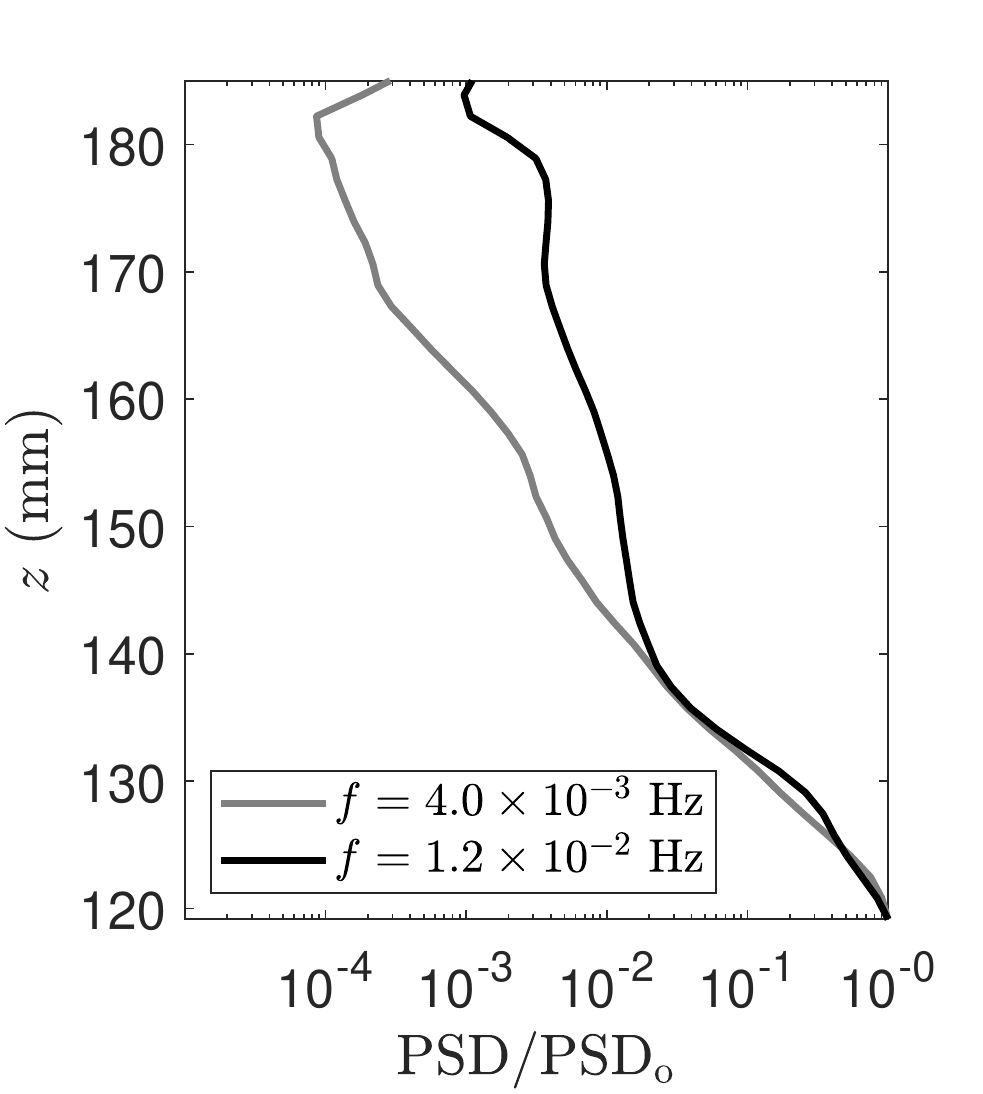}
    \caption{(Left) Power spectral density of the absolute velocity $\sqrt{u^2+w^2}$ above the convective layer. The grey curve shows the theoretical buoyancy frequency profile, assuming a linear profile for the temperature, from $4^\circ$C to $35^\circ$C. (Right) Two selected profiles (taken at frequencies shown by dashed lines on the left graph) of the re-scaled PSDs by the PSD at the top of the convective layer, \textit{i.e.} $z=120$~mm (noted $\mathrm{PSD_o}$). The PIV measurements are performed for 50 minutes and results (using the pwelch Matlab function) are horizontally averaged at each height to obtain the averaged power spectral density.}
    \label{fig:spectro}
\end{figure*}

To further investigate the waves signal, waves spectra are plotted in figure \ref{fig:spectro}, showing the power spectral density of oscillatory motions within the stratified layer at every height, averaged horizontally for each height. The grey line is the theoretically computed buoyancy frequency profile. Figure \ref{fig:spectro} shows that energy is present in a wide frequency band, from the lowest measured frequencies to the buoyancy frequency $N$. Low frequency motions $f < 4\times10^{-3}$~Hz are very intense and propagate high in the stratified layer. Motions with frequency ranging from $4 \times 10^{-3}$~Hz to $N$ are less intense, but still propagate into the stratified layer. Motions propagating at frequencies higher than the buoyancy frequency $N$ are greatly attenuated after a centimetre as IGWs of frequency larger than $N$ are evanescent. The weak signal at low frequencies above $z=180$ mm comes from the convective motions due to the non-homogeneous heating at the top. These motions are confined at the very top of the experimental container. 

Right panel of figure \ref{fig:spectro} shows two vertical profiles of the PSD re-scaled by the PSD at the top of the convective layer ($z=120$~mm), taken at two different frequencies. The energy decrease is quite similar between $z=120$~mm and $z=140$~mm for both frequencies. However, for $z>140$~mm, the energy for the higher frequency decreases slower than the energy for the lower frequency. This dependence of the attenuation length regarding the frequency of the signal is a characteristic of IGWs. Indeed, the dispersion relation of IGWs relates the frequency and the wave vector direction. Moreover, energy propagates perpendicularly to the wave vector for IGWs (i.e. group and phase velocities are perpendicular). The closer to $N$ the wave frequency $\omega$ is, the more horizontal the phase propagates, hence the more vertical energy propagates. High frequency waves are thus capable of transporting energy to high altitudes before being damped. On the opposite, waves with low frequency compared to $N$ propagate energy almost horizontally, and are thus attenuated before reaching high altitudes. At frequencies $f<4\times 10^{-3}$~Hz, a lot of energy is seen and the attenuation length does not depend on the frequency. There is no reason why IGWs should disappear below a certain frequency, but we would expect to see the attenuation length to keep decreasing with decreasing frequency. We thus deduce that IGWs at frequencies $f \leqslant 4\times 10^{-3}$~Hz are hidden in the energy spectrum by some very energetic large-scale slowly-varying flow, which we will describe below.

More than one order of magnitude separates the buoyancy frequency and the fastest large-scale flow fluctuations. The large-scale flow penetrates deep into the stratified layer. It globally decreases in amplitude with height, but with some local increases at $z\sim 125$ mm (i.e. close to the interface between convective and buffer layers) and $z \sim 145$ mm.
The IGWs signal can be seen between $f = 4 \times 10^{-3}$ Hz and the buoyancy frequency. A peak that reaches the top of the stratified layer is seen around $f= 1.2 \times 10^{-2}$ Hz, \textit{i.e.} the same frequency as the convective forcing discussed in section \ref{sec:results_conv}. It corresponds to the strong excitation provided by the cold rising and hot sinking turbulent plumes.
However, top panel of figure \ref{fig:spectro} also shows a sudden drop of the energy at frequencies $f>1.2 \times 10^{-2}$~Hz. Indeed, wave attenuation is strong at these frequencies, even if they are close to (but below) the buoyancy frequency $N$. Actually, energy dissipation also depends on the norm of the wave vector squared. There is no reason that all excited waves have the same wave vector norm; one could even expect that fastest waves are excited by fastest, hence smallest convective patterns, and are thus also at smallest scale: they then dissipate more rapidly.

\subsubsection{\label{sec:results_lsf}Large-scale flow in the stratified layer}
Figure \ref{fig:spectro} shows an important amount of energy at low frequencies which has been interpreted as the signature of a large-scale slowly-varying flow in the stratified layer. We will now investigate the nature of these fluctuations to see if they relate to reversals similar to the QBO.

\begin{figure*}[t]
    \centering
    \includegraphics[scale = .8]{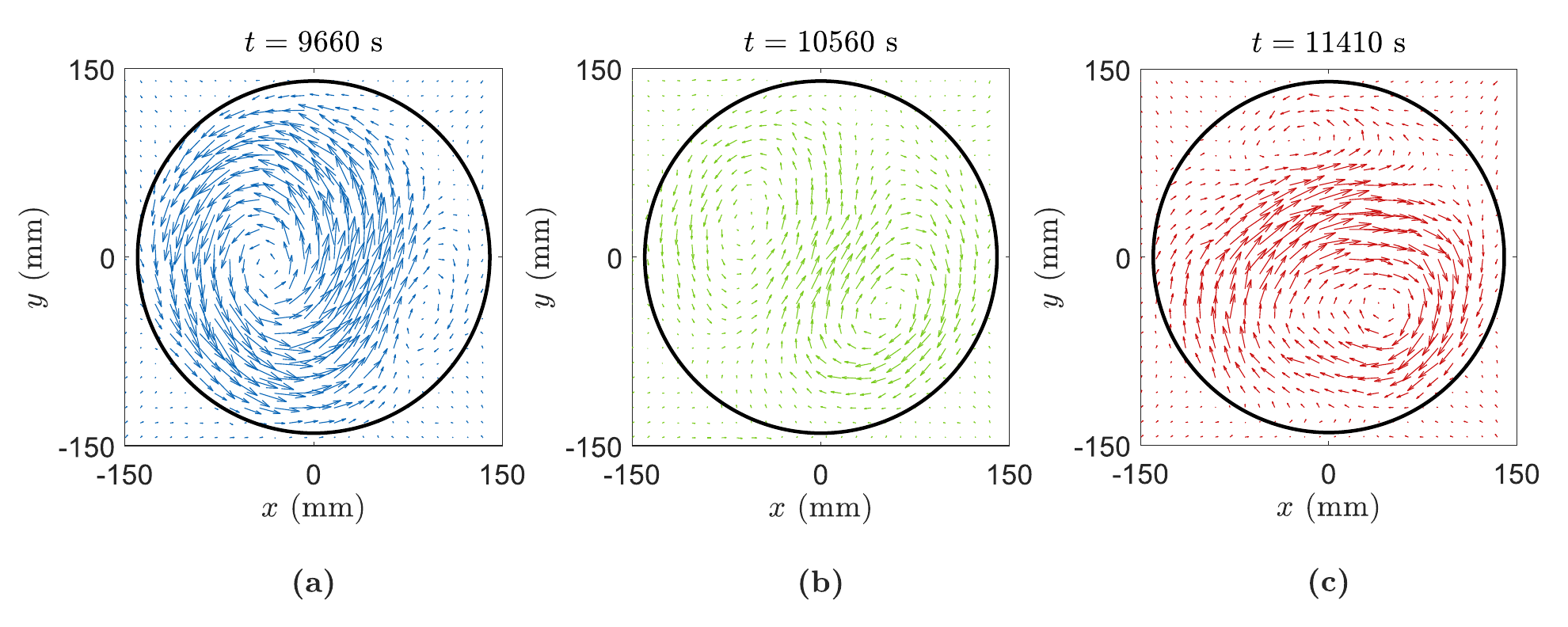}
    \caption{Horizontal velocity vector fields in the stratified layer at different times. The laser sheet is located at $z=150$~mm. The large-scale flow reverses from (a) to (c). Time between (a) and (c) is approximately half an hour. Maximum velocities are $0.1$ mm/s.}
    \label{fig:meanflow}
\end{figure*}

\begin{figure}[t]
    \centering
    \includegraphics[scale = .45]{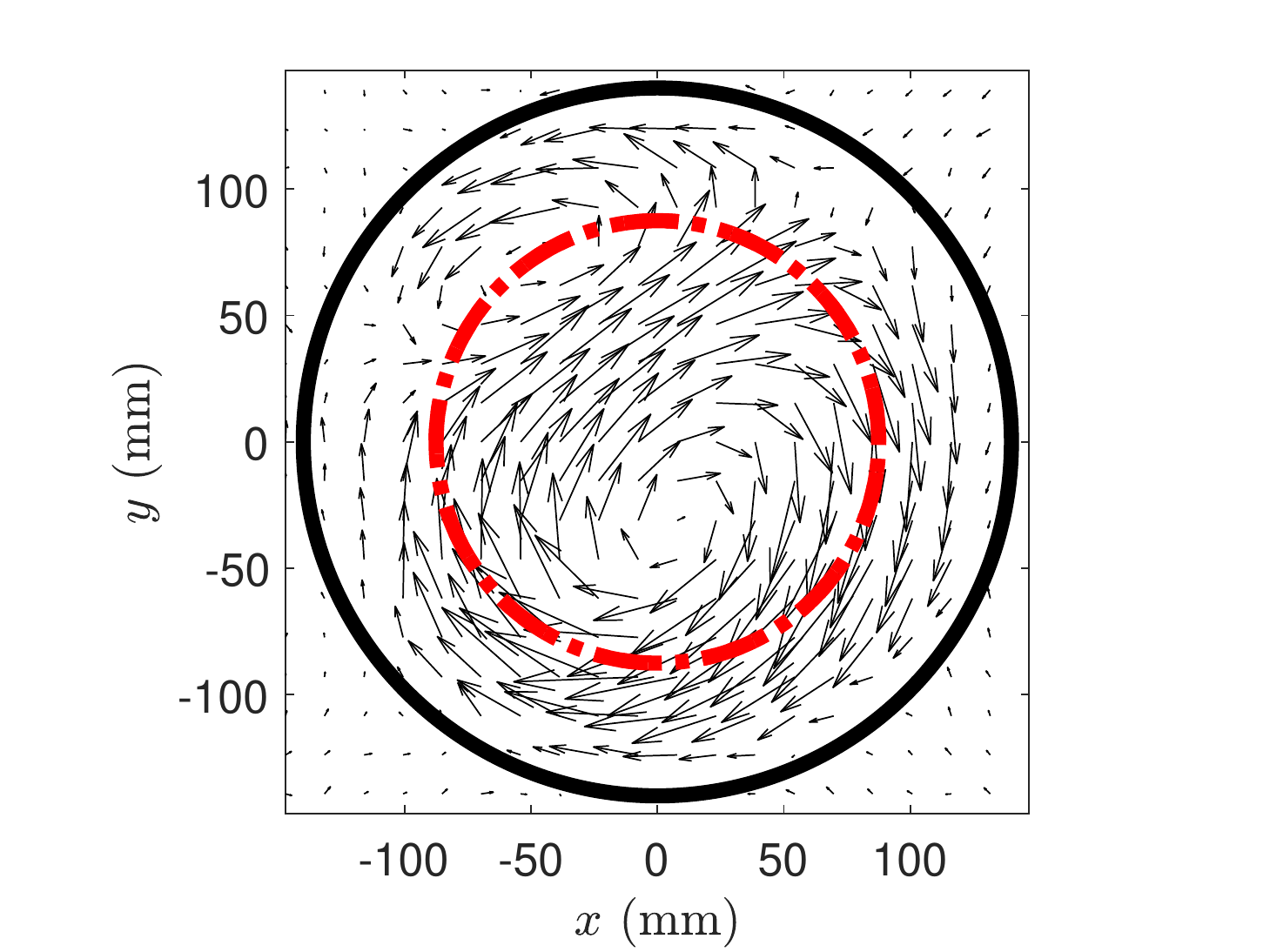}
    \caption{Criterion used to extract a significant value for the large scale flow and its direction: the azimuthal velocity is averaged over the ring shown in red.}
    \label{fig:critere}
\end{figure}

Figure \ref{fig:meanflow} shows horizontal vector fields at the same depth at different times. In figure \ref{fig:meanflow}(a), the flow goes counter-clockwise inside the cylinder. Figure \ref{fig:meanflow}(b) shows that two contra-rotating vortices with a smaller amplitude typical velocities have appeared. Figure \ref{fig:meanflow}(c) shows a mostly clockwise rotating flow, where one of the preceding eddy pairs has nearly disappeared. The large-scale flow thus evolves drastically over time. A criterion is computed to extract a typical mean velocity from those fields that accounts for the ``direction'' of the large-scale flow: as illustrated in figure \ref{fig:critere}, we compute a mean azimuthal velocity, taken along a ring centred in the cylinder. Other criteria to extract a representative value for the large scale flow direction have also been tested, including: the mean vorticity over the cylinder area, the average of the azimuthal velocity over several rings with different radii, and the azimuthal velocity averaged over thick rings. They all give similar results for the large-scale flow measurement. 
\\

In order to investigate the vertical phase propagation of the reversals, and thus, to compare the reversal dynamics observed to a QBO-like phenomenon, the setup has been equipped with a linear translating platform that allows us to perform horizontal laser sheet sweeping along the vertical. Horizontal velocities are measured in an horizontal plane, every $5$~mm from the top of the convective layer $z=110$~mm to the middle of the stratified layer $z=160$~mm. Any trace of downward phase propagation of the reversals, as observed on the QBO on Earth \cite{baldwin_quasi-biennial_2001} and on the historical Plumb experiment \cite{plumb_interaction_1977, semin_nonlinear_2018}, would be a significant evidence for QBO-like phenomenon in the experiment. 
Indeed, the phase propagation of the reversals due to IGWs non-linear interactions is theorised as follows: an IGW propagating in a stratified layer with an horizontal phase velocity in the same direction as the existing base flow propagates upward until reaching a critical height $z_c$, where it deposits all its energy locally. At $z=z_c$, the flow accelerates. Thus the critical height where the flow is intense enough to damp the wave is lowered. As time goes on, this critical height moves towards the location where the waves are emitted. Here, the waves are emitted at the bottom of the stratified layer. We would expect a downward phase propagation if the reversals are driven by IGWs non-linear interactions.

\begin{figure*}
    \centering
    \includegraphics[scale=0.7]{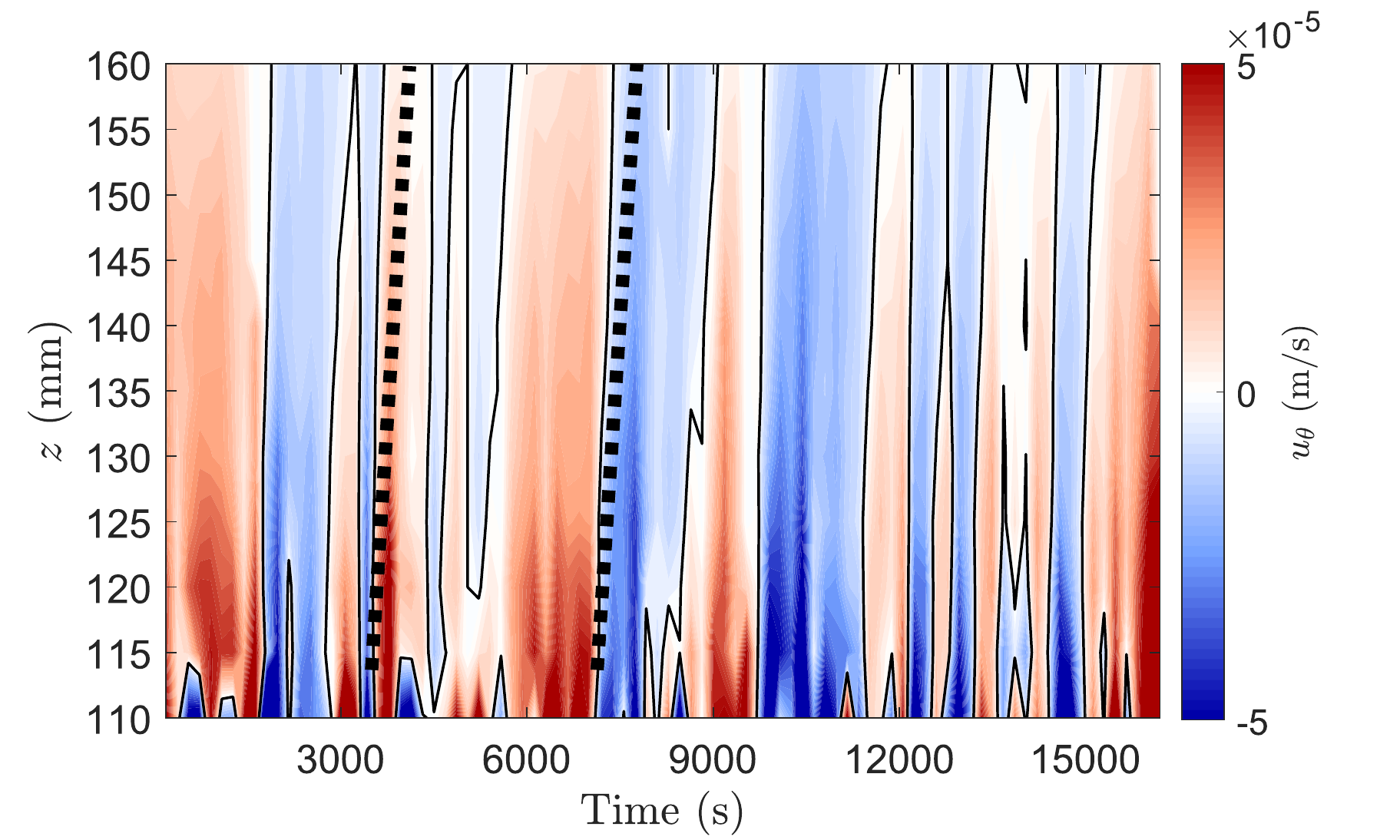}
    \caption{Reversals of the large-scale flow. $z=115 - 120$ mm is the convective / buffer layers interface. Ascending plumes often perturb the buffer layer flow. The velocity was measured at 11 heights, marked by each tick in the vertical axis of the figure, and interpolated in between. The slope of the black dot lines represent the viscous coupling phase velocity.}  \label{fig:QBO}
\end{figure*}

We performed long time experiments (around 8 hours). Typical results extracted from the criterion described above are shown in figure \ref{fig:QBO}. Blue patches (resp. red patches) represent large scale flow going counter clockwise (resp. clockwise). The present measurements mainly confirm the interpretation of figure \ref{fig:spectro} for the lowest frequencies: the large-scale flow is horizontal and extends over the whole depth of the stratified layer with an amplitude attenuated with height, and exhibits slow reversals. Additionally, some intense events at $z = 110$ mm are directly related to penetrative plumes from the convection. Reversals times range from $400$s to $1800$s. However, no downward phase propagation of the reversals is observed. On the contrary, the reversals seem to occur along the whole stratification height at the same time, or even with a rapid upward phase propagation. Since the phase propagation is not towards the location where the waves are emitted, the reversals are unlikely driven by the non-linear interactions of IGWs. However, as seen in section \ref{sec:results_waves}, IGWs propagate in the stratified layer and carry energy. Therefore, they give energy to the large-scale flow through non-linear interactions. Yet, the process is not dominant in the reversals dynamics.
\\
Since, the reversals observed in figure \ref{fig:QBO} do not have a downward phase propagation, we look for other mechanisms than the QBO mechanism to explain the reversals. Two other mechanisms can be investigated. The first one relies on a specific convective dynamics within the overall stratified layer driven by horizontal gradients related to imperfect top and side boundary conditions. The second mechanism relies on viscous coupling with the underlying convective and buffer layers. 

Our fully stratified reference experiment described in section \ref{sec:methods} precludes the first scenario. Indeed, setting the bottom boundary at 10$^{\circ}$C and the top boundary at 70$^{\circ}$C, no motion is observed for the bottom $3/4$ of the tank. In this test-experiment, the top $1/4$ of the tank is animated by convective motions due to the non-homogeneous top heat source (in the standard $4^\circ$C experiment, where $T_{top} = 35\degree$C, only $\sim 2$ cm are affected by the convection at the top of the tank, because the non-homogeneity of the heat source is less important for lower temperature, thus the horizontal convection is weaker). However, these are inefficient to generate waves below and to drive any large-scale flow observable away from the top region.

\begin{figure}[h]
\includegraphics[scale = 0.55]{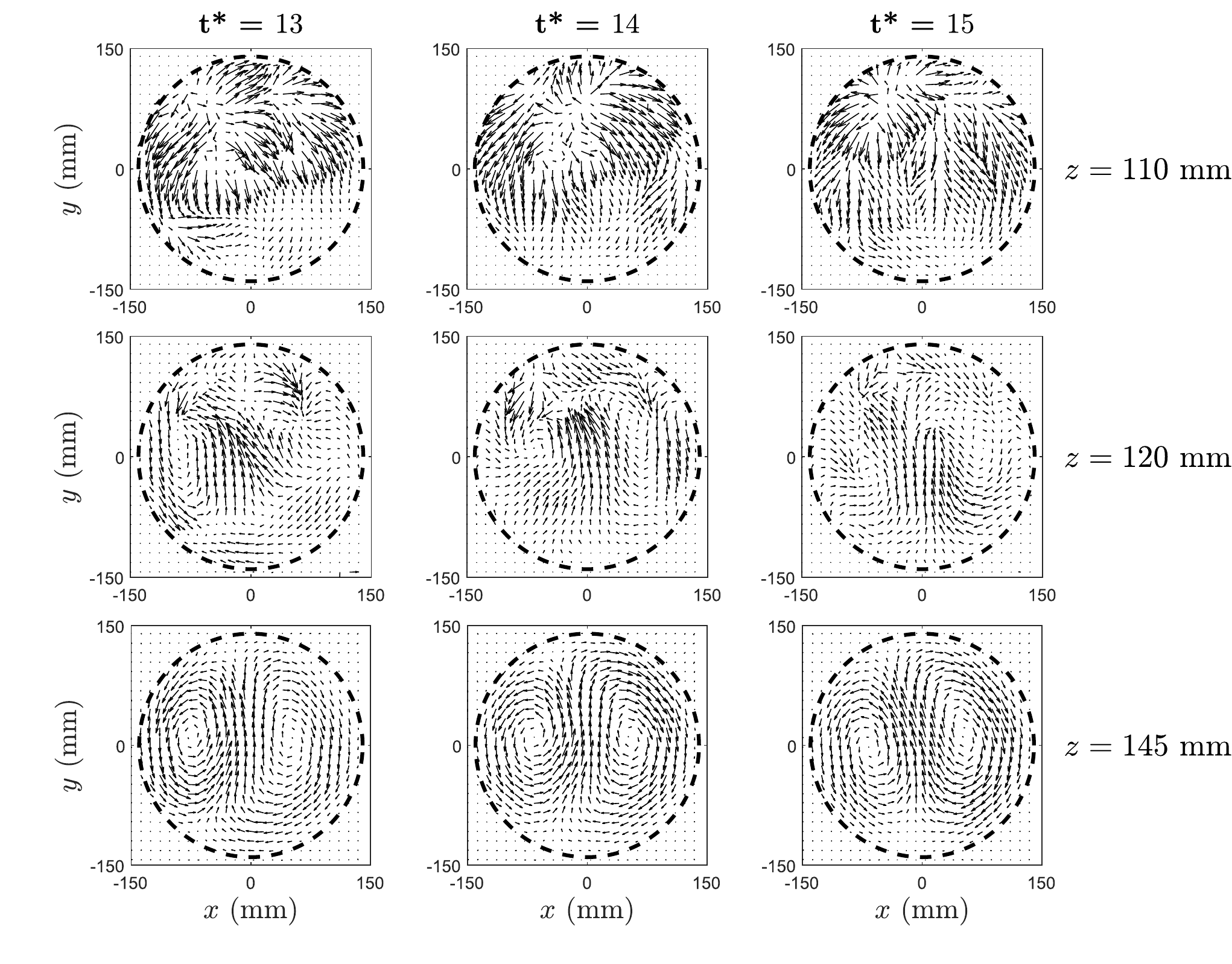}
  \caption{Velocity vector fields in the horizontal plane. Different columns represent different sweeping cycles $t^*$ (one sweeping cycle corresponds to the 11 steps needed to go from the lowest position $z=110$~mm to the highest position $z=160$~mm). Different rows represent different heights within the same sweeping cycle: first row is the top of the convection $z=110$~mm, second row is in the buffer layer $z=120$~mm and third row is in the stratified layer $z=145$~mm. Convective plumes are easily noticeable on the first row fields.}
  \label{fig:champ_correle}
\end{figure}

\begin{figure}[h]
  \includegraphics[scale = 0.55]{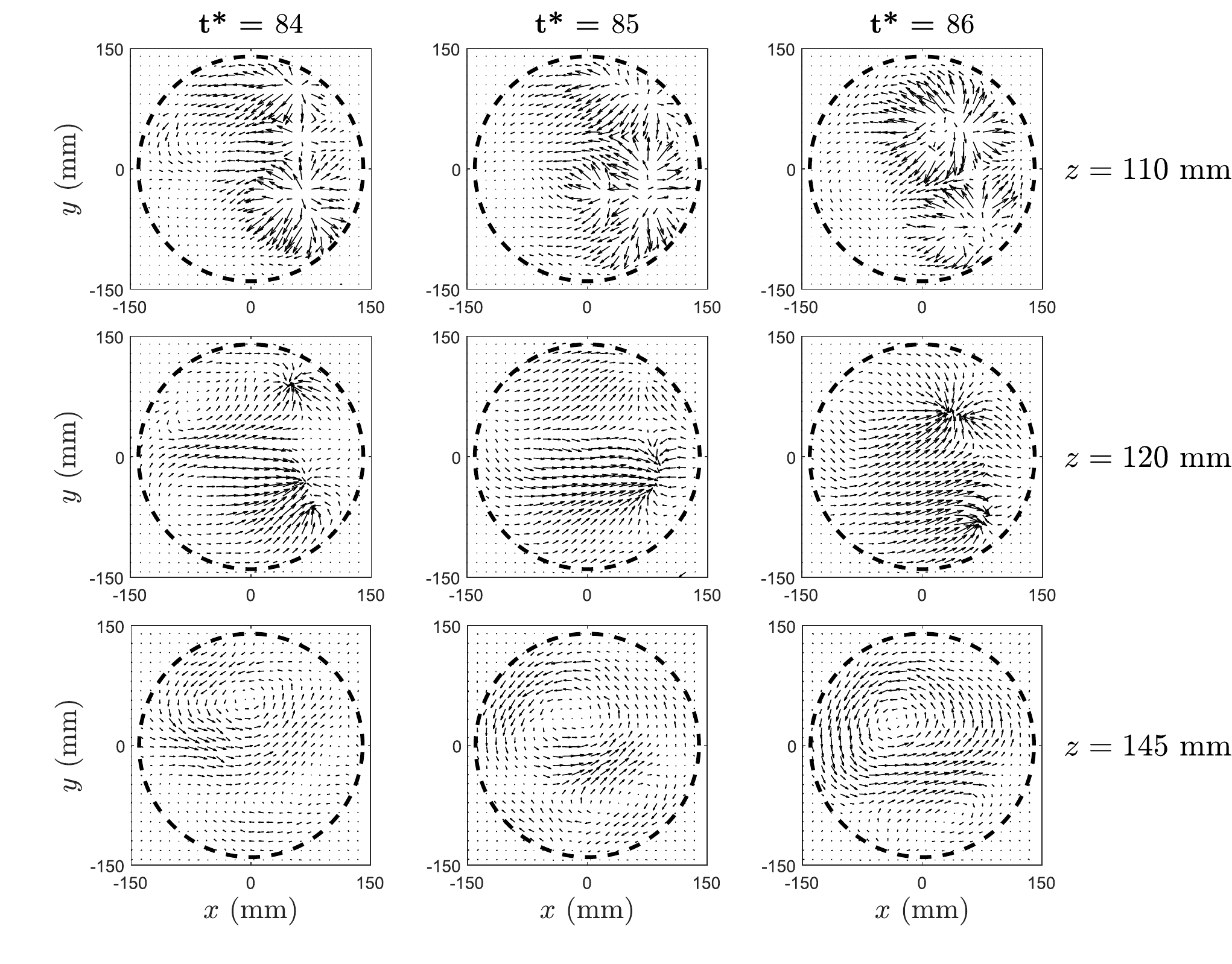}
  \caption{Same as figure \ref{fig:champ_correle} but for different sweeping cycles. Note that in this set of velocity fields, the buffer and stratified layers are less correlated than they are in figure \ref{fig:champ_correle}.}
  \label{fig:champ_pascorrele}
\end{figure}

\begin{figure}[h]
  \centering
  \includegraphics[scale=0.6]{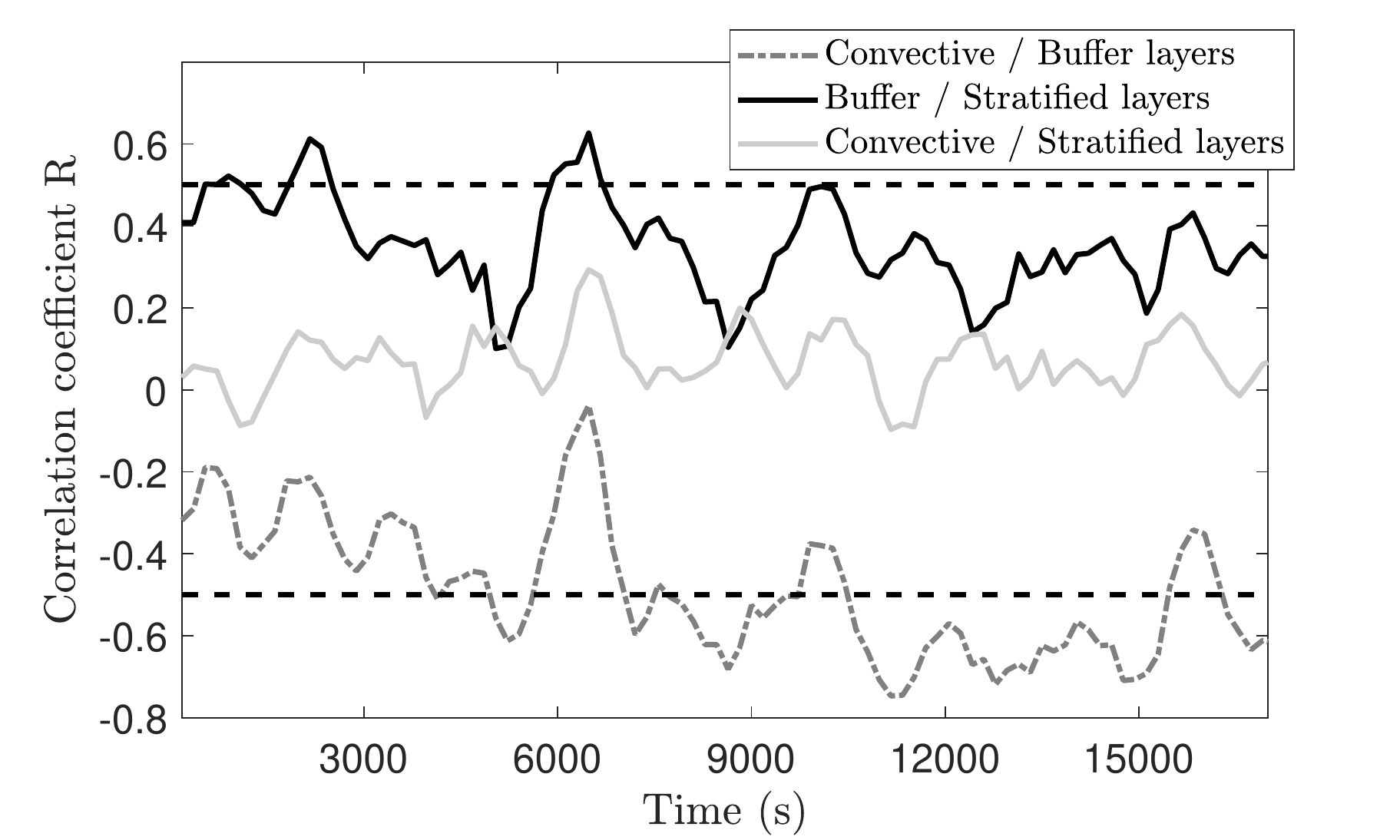}
  
  \caption{Velocity correlation between the three layers. Dashed black lines are the $-0.5$ and $0.5$ values. Correlation coefficient are  window-averaged over $10$~min to smooth the curve.}
  \label{fig:correlation}
\end{figure}

This leaves viscous entrainment as a possible driving mechanism. The dotted lines on figure \ref{fig:QBO} show a theoretical viscous time, computed from the time for viscous entrainment to drive 20\% of the horizontal velocity at $z = 115$ mm to $z=160$ mm, starting from a base state flow at rest. The 20\% value corresponds to the measured value of the large scale flow at $z=160$ mm compared to the value at $z = 115$ mm (noted $u_b$). The theoretical viscous time entrainment is given by $t = \frac{z^2}{4\,\nu\, \mathrm{erf}^{-2}\left(\left(\frac{u}{u_b}-1\right)\right)}$. The reversals occur in a time scale comparable with this theoretical viscous time. The similarity between the slope of the dashed lines and the slope of the upward phases suggests that reversals are driven viscously.

However, the existence of the buffer layer and its associated intense shear, with opposite horizontal velocities with the convective layer below (see figure \ref{fig:bufferlayer}) precludes direct viscous coupling between the convective and stratified layers. Besides, no reversal has been observed in the convective region. We thus propose a thermal coupling between the convective and buffer layers as seen in section \ref{sec:results:buffer}, associated to a viscous coupling between the buffer and stratified ones. To further quantify this possibility, figures \ref{fig:champ_correle} and \ref{fig:champ_pascorrele} show horizontal velocity fields at different heights and at different times. For each of the columns shown, first row is the mean flow in the convective layer at depth $z=110$~mm, second row is the mean flow in the buffer layer at depth $z=120$~mm and third row is the mean flow in the stratified layer at depth $z=145$~mm. The correlation coefficients through time between (i) the convective and buffer layers, (ii) the buffer and stratified layers, and (iii) the convective and stratified layers have been computed. It consists in a scalar product of the velocity vector for each position at two different heights rescaled by the product of the norm of the velocity vector at the two heights, \textit{i.e.}: 

\begin{equation}
R_{ij} = R(x_i,y_j) = \frac{u(x_i,y_j,z_1) \times u(x_i,y_j,z_2) + v(x_i,y_j,z_1) \times v(x_i,y_j,z_2)}{\left(u(x_i,y_j,z_1)^2+v(x_i,y_j,z_1)^2\right)^{1/2} \times \left(u(x_i,y_j,z_2)^2 + v(x_i,y_j,z_2)^2 \right)^{1/2}}
\end{equation}
This gives a correlation coefficient $R_{ij}$ for each PIV position in the horizontal plane. The global correlation coefficient $R$ is computed by spatially averaging the local correlation coefficients. 

Results are shown in figure \ref{fig:correlation}. The convective and buffer layers are negatively correlated: the correlation coefficient is most of the time close to $R=-0.5$. This can also be seen at all times in figures \ref{fig:champ_correle} and \ref{fig:champ_pascorrele}, where horizontal velocities in the convective and buffer layers have opposite direction. A diverging flow coming from an impinging plume in the convective zone corresponds to a converging flow in the buffer layer towards the impact zone, hence confirming the thermal coupling mechanism described in section \ref{sec:results:buffer}. This converging flow may lead either to a clockwise or anticlockwise azimuthal mean flow, depending on the details of the chaotic excitation from the convective plumes. The correlation coefficient between the convective and stratified layers can be positive or negative, and is anyway most of the time less than 0.2, in absolute value. The correlation coefficient between the buffer and stratified layers shows a lot of temporal variations. However, it remains always positive. At a given time, the large-scale flow in the stratified layer may switch between a regime strongly dominated by the buffer layer (see also figure \ref{fig:champ_correle}), and a second regime where the flow in the stratified layer is quite different from the flow in the buffer layer (see also figure \ref{fig:champ_pascorrele}). 

We thus conclude that the stratified layer is globally viscously driven by the buffer layer. However, the stratified layer exhibits additional complexities. These might be due to IGWs interacting with the large-scale flow. The results from Couston et al. \cite{couston_order_2018} show that the lower the Prandtl number, the more regular the QBO. In the experiment, the Prandtl number is close to $Pr = 7$: the typical associated QBO-type flow is irregular, with low amplitude. We thus propose that large-scale flow driven by IGWs non-linear interaction superimposes on the viscously driven flow, but remains secondary. We do not know at this point how to disentangle those two potential contributions from the available data. 


\clearpage
\subsection{\label{sec:results_num} Numerical simulations}
The experimental results are not fully sufficient to explain, with complete certainty, the origin of the buffer layer and of the large-scale flow observed in the stratified layer. In addition, the effects of the lateral heat losses and top temperature heterogeneity are difficult to distinguish. To answer these questions, 3D DNS of a configuration similar to our experiments are performed, reproducing the 4$^{\circ}$C convection but with idealised boundary conditions (i.e. no flux on the sides, and fixed temperature at the top and bottom). As mentioned in section \ref{sec:methods_num}, the Rayleigh number $Ra$ and $T_{top}$ are tuned so that the interface depth in the experiment and the numerical simulation are similar. We have $Ra=10^7$ and $T_{top}=48^\circ$C. All the numerical simulations are run dimensionless, but results are shown in dimensional values. The length scale is $H = 200$~mm, the vertical extent of the whole domain (hence diameter is $D=300$~mm), the timescale is the thermal diffusive time $\tau = \frac{H^2}{\kappa} = \frac{0.2^2}{1.5 \, 10^{-7}} = 2.67 \times 10^{5}$~s, and the temperature is given by the dimensionless temperature $\theta = \frac{T - T_i}{T_0 - T_i}$, where $T, T_i, T_0$ are respectively the dimensional temperature, the inversion temperature of the equation of state (i.e. $4^\circ$C), and the bottom temperature (i.e. $0^\circ$C). Results for sections \ref{sec:results_num_conv} - \ref{sec:results_num_igw} are computed from a $(x,z)$ vertical plane located along a cylinder diameter.

\subsubsection{\label{sec:results_num_conv}Large-scale circulation in the convection zone and buffer layer}
\begin{figure}[h]
    \centering
    \includegraphics[scale = .47]{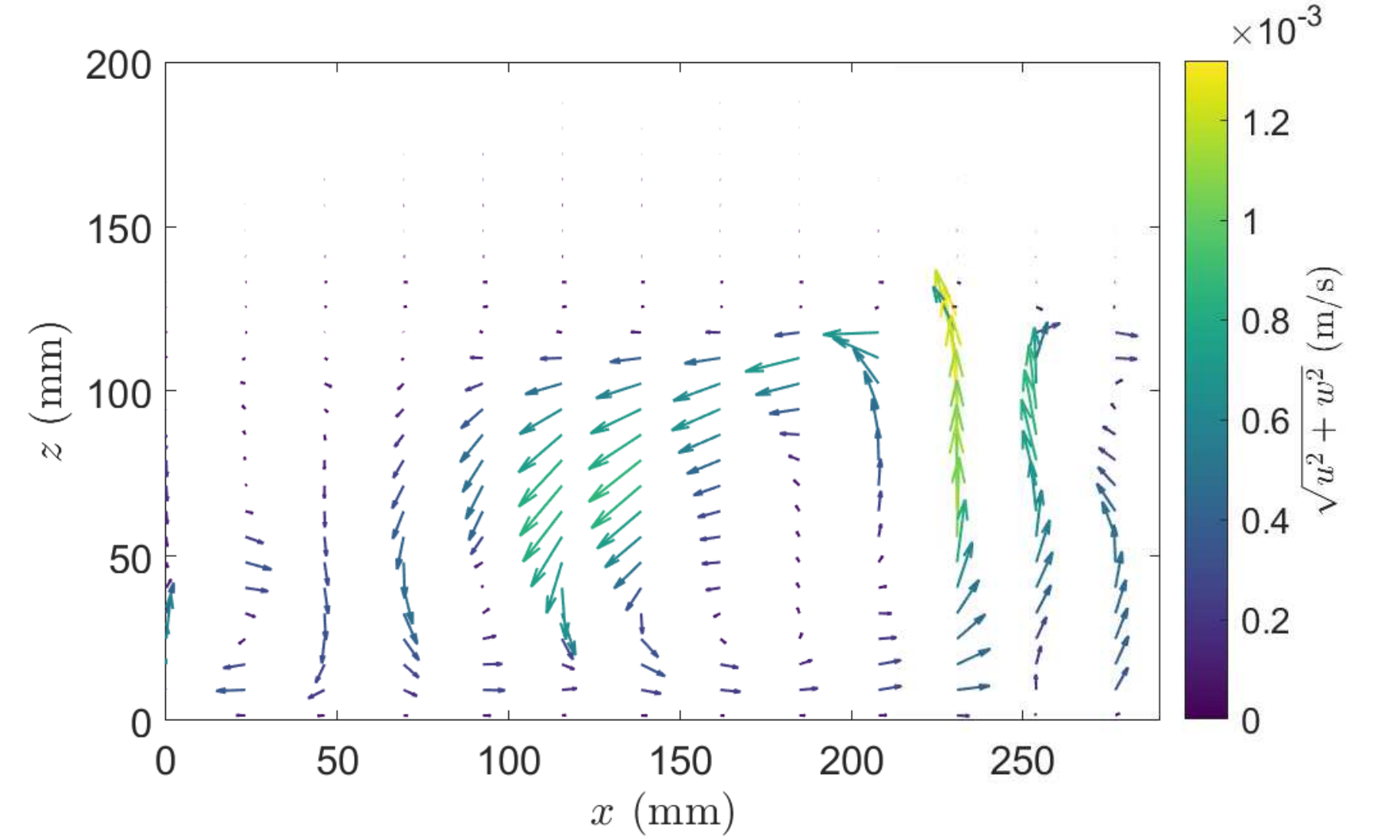}
    \hspace{.1cm}
    \includegraphics[scale = .47]{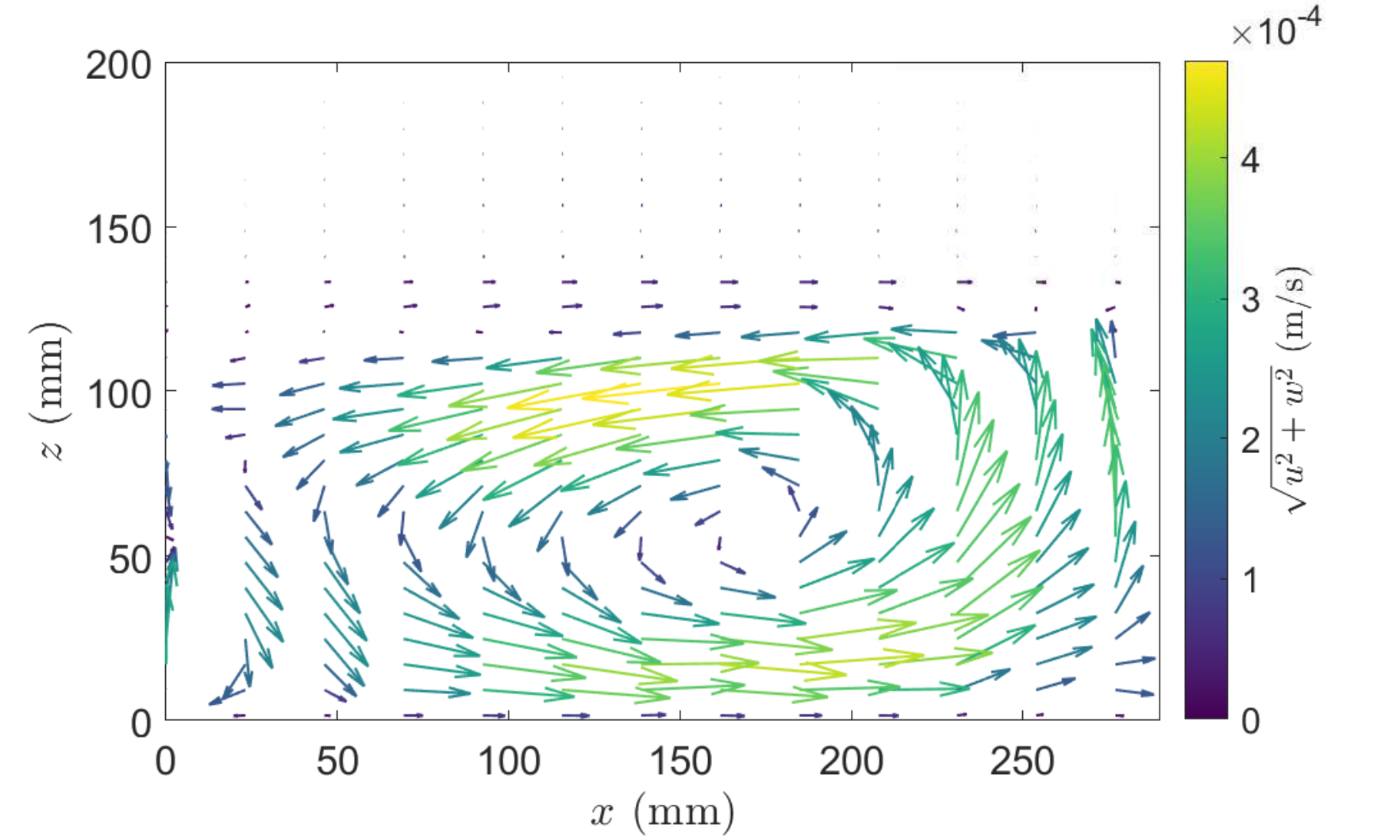}
    \caption{(Left) Instantaneous velocity field. An ascending plume is visible at $x=230$ mm. (Right) Large-scale circulation in the convective layer obtained by time-averaging velocities over a 50 minutes recording. The large-scale circulation is a counter clockwise cell. Maximum instantaneous velocities are $3$ times bigger than the maximum averaged velocities.}
    \label{fig:LSC_num}
\end{figure}

Figure \ref{fig:LSC_num} shows that a large-scale circulation takes place in the convective layer. It consists of a cell filling the whole convective layer, and exhibits no reversal over the whole course of the simulation. The fluid rotates counter clockwise in the vertical plane. This is qualitatively consistent with the mean flow observed in the experiment and shown in the right panel of figure \ref{fig:LSCexpe}. As in the experiments, a counter current exists on the top of the convective layer at $z= 120$~mm, creating a strong shear and demonstrating the existence of a buffer layer in the numerical simulation as well. 

\begin{figure*}[t]
    \centering
    \includegraphics[scale = 0.5]{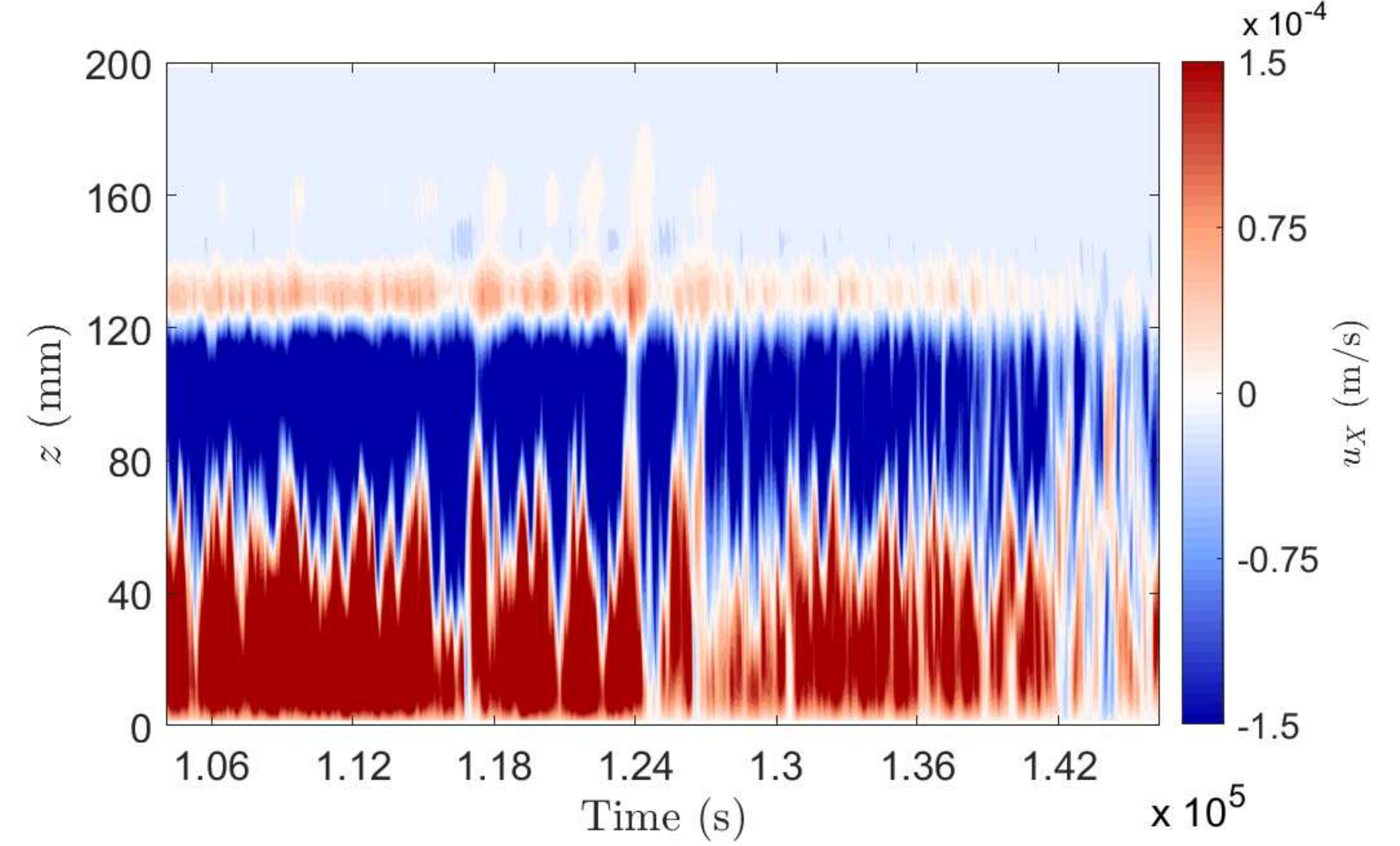}
    \hspace{.1cm}
    \includegraphics[scale = 0.5]{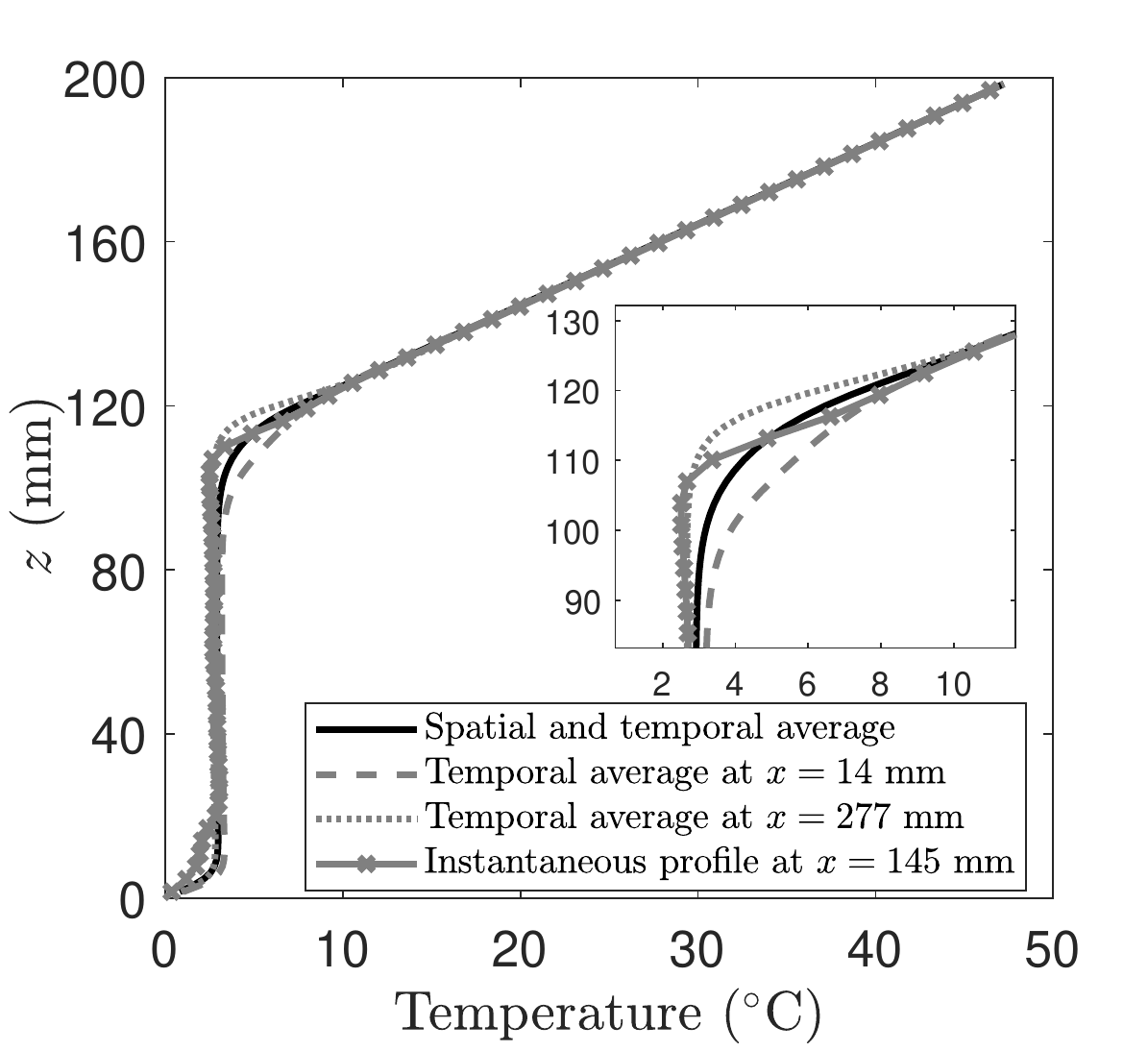}
    \caption{(Left) Horizontal average of the horizontal velocity $u$ over a vertical cross-section in the middle of the tank. The buffer layer can be seen above $z=120$~mm. A stationary large-scale circulation is present in the convective layer, even if it appears quite perturbed at the end of the signal. (Right) Temperature profiles along the $z$-axis.}
    \label{fig:num_buffer}
\end{figure*}

The space-time diagram of the mean horizontal flow shown in figure \ref{fig:num_buffer} confirms it. Observing the buffer layer in the absence of side thermal losses and top temperature heterogeneity is an additional argument accounting for the fact that it is not an artefact driven by imperfect experimental conditions.
We also observe that the flow within the convection stays positive through time at the bottom and negative at the top. This is evidence of the steady large scale circulation taking place in the convective layer. Some events appear at $t > 1.42 \times 10^{5}$~s and are interpreted as quasi-reversal of the large-scale circulation.

\begin{figure}[t]
    \centering
    \includegraphics[scale=.5]{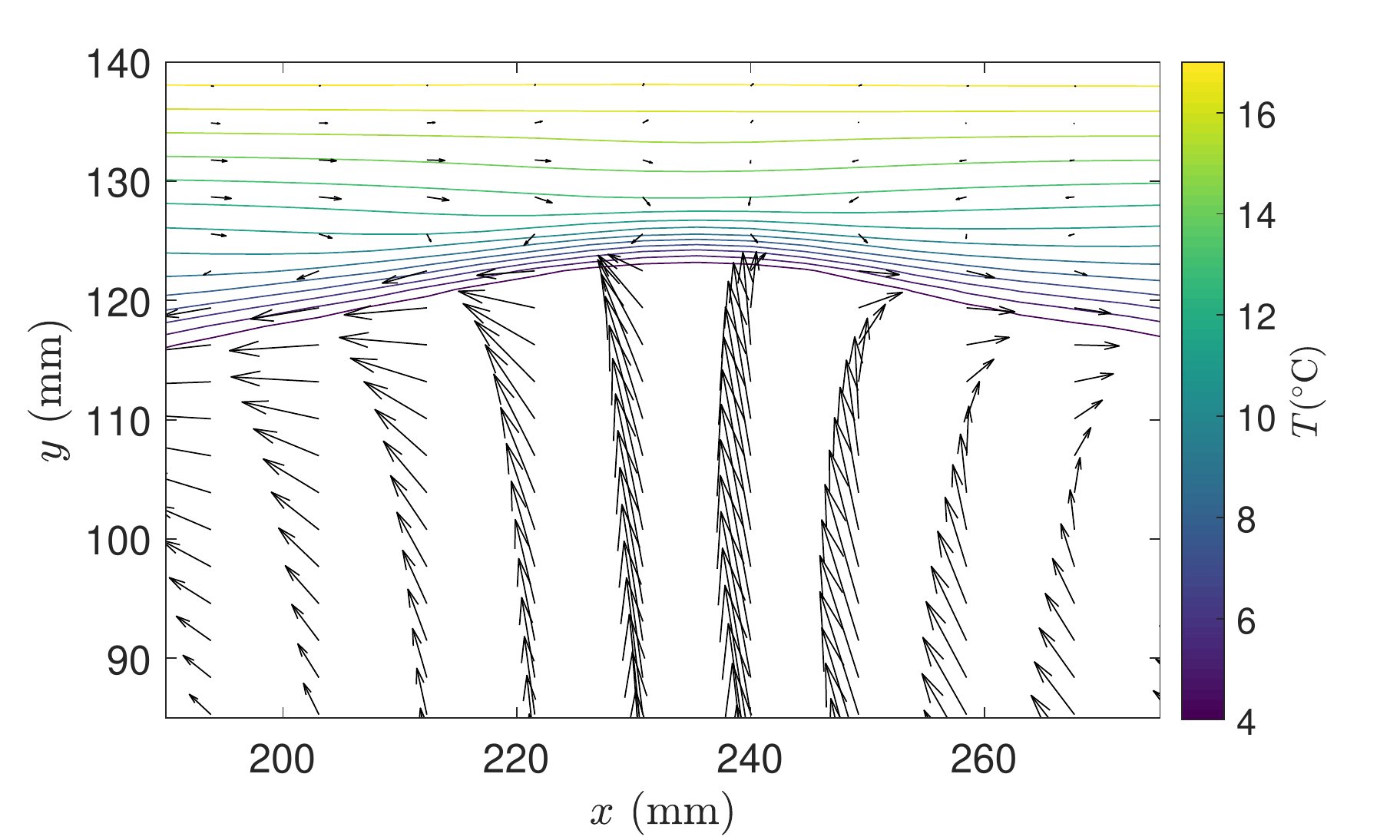}
    \caption{Velocity field and temperature isotherms at the end of an upward plume impact on the interface.}
    \label{fig:num_quiverisoT}
\end{figure}

\begin{figure}[t]
    \centering
    \includegraphics[scale=.6]{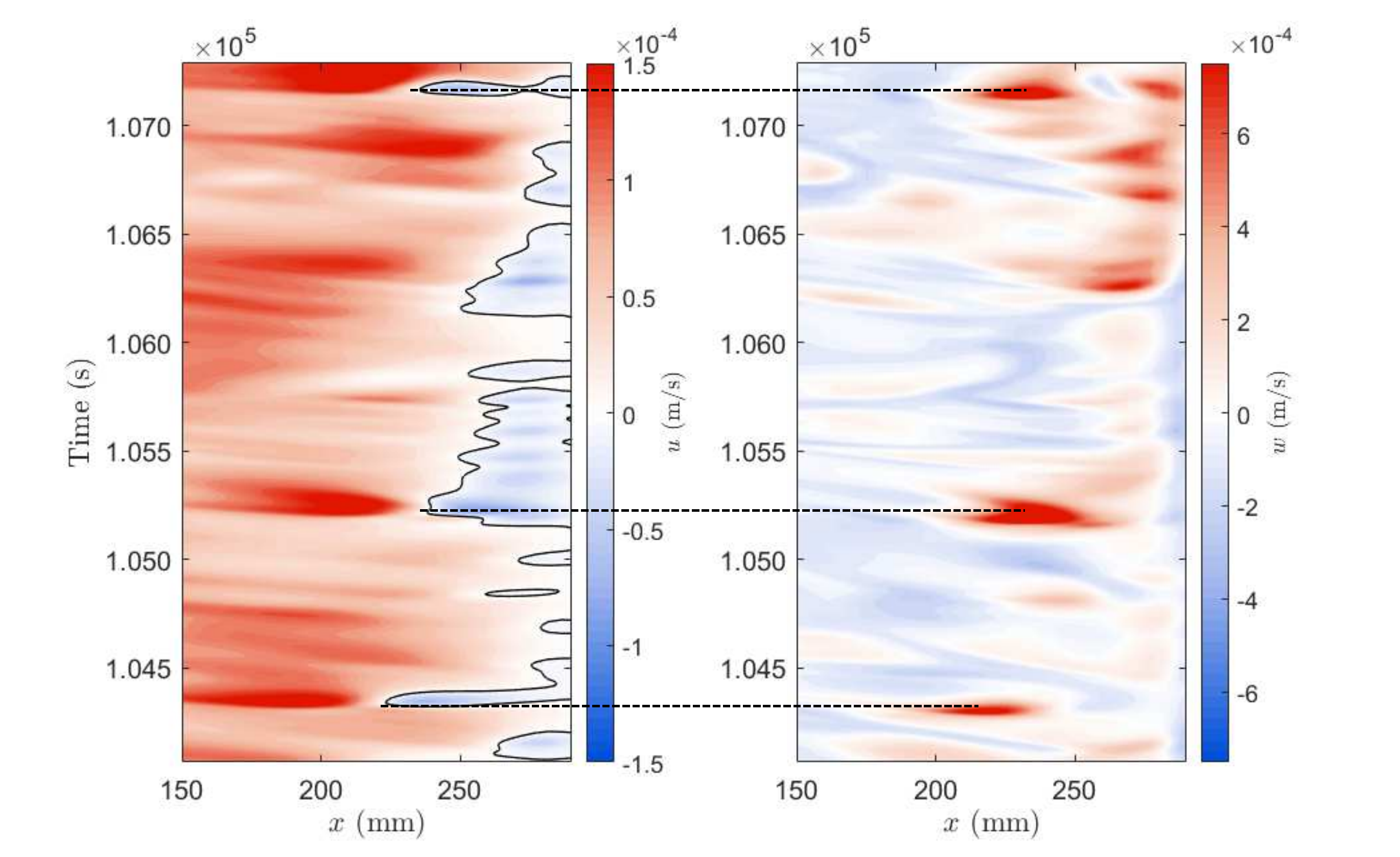}
    \caption{(Left) Spatio-temporal diagram of the horizontal velocity $u$ at $z=128$~mm. (Right) Spatio-temporal diagram of the vertical velocity $w$ at $z=108$~mm. The event at $t \approx 1.052 \times 10^5$~s is shown in figure \ref{fig:num_quiverisoT}.}
    \label{fig:num_correWU}
\end{figure}

The temperature profile along the $z$ axis is also plotted on the right panel of figure \ref{fig:num_buffer}. The figure shows a temporal and horizontal average of the temperature field (black thick curve), two temporal averages at two different positions $x=14$~mm (left side of the tank - dashed grey) and $x=277$~mm (right side - dotted grey) and an instantaneous profile at $x=145$~mm (middle of the tank - thick grey with crosses). The thermal boundary layer can be seen, between $z=0$~mm and $z=10$~mm. Then, between $z=10$~mm and $z=100$~mm lies a layer of constant temperature $T \sim 2.8^\circ$C. Between $100 \mathrm{~mm} \leqslant z \leqslant 115 \mathrm{~mm}$, the temperature profile evolves from constant to linear for $z > 115$ ~mm. The $T=4^\circ$C (respectively $T=8^\circ$C) isotherm is located at $z = 110$~mm (resp. $z = 120$~mm). Note that the temporal average of the temperature profiles are different on the left and right sides of the tank. Indeed, the constant temperature height goes to $z=90$~mm for the left side whereas it goes to $z=115$~mm for the right side. This suggests that the convective / buffer layer interface does not lies at one height over the whole tank but is a function of time and space. This is very likely due to the large-scale circulation. Thus, the thermal coupling described in \ref{sec:results:buffer} will likely occur at different heights, depending on time and horizontal position.

The thermal coupling as schematised in figure \ref{fig:schema_couplage} can be found in the numerical simulation. This is represented in figure \ref{fig:num_quiverisoT}. An upward plume impacting the convective / buffer layer interface is seen. The isotherms ranging from $T=4^\circ$C to $T=11^\circ$C are deflected upward, due to the plume bringing cold fluid upward. On the contrary, the isotherms $T = 12 -14^\circ$C are deflected downward by the converging flow. Isotherms at $T \geqslant 15^\circ$C remain horizontal. After the impact on the interface, the plume is deflected outwards. One could expect the fluid above the impact to be viscously entrained by this outward deflection. However, as observed in figure \ref{fig:num_quiverisoT} for the simulation and figures \ref{fig:champ_correle}-\ref{fig:champ_pascorrele} for the experiment, the fluid above the interface is going towards the plume, \textit{i.e.} in the opposite direction of the fluid below, hence explaining the observed shear (see figures \ref{fig:num_quiverisoT} and \ref{fig:schema_couplage}). The time evolution of these dynamics is shown in figure \ref{fig:num_correWU}.

Figure \ref{fig:num_correWU} shows the time evolution of the horizontal velocity $u$ in the shear layer at $z=128$~mm and the time evolution of the vertical velocity $w$ in the convective layer at $z=108$~mm. Comparing the two panels of figure \ref{fig:num_correWU} shows that upward plumes are concomitant with converging horizontal velocities towards the plume impact. Indeed, the spatio-temporal diagram of $w$ exhibits local strong upward plumes. These plumes, as suggested by the dashed black lines, are correlated in time and space with converging horizontal velocities. For instance, an upward plume is seen at $x\approx220$~mm and $t\approx1.043 \times 10^5$~s. At the same horizontal position and time, the positive horizontal velocity becomes stronger and the negative horizontal velocity patch increases in size to reach $x\approx220$~mm. The converging horizontal velocities event occurs a short time after the impact of the plumes. Thus, it can be concluded that the plume induces the converging flow, as suggested by our explanation in section \ref{sec:results:buffer}.

\clearpage

\subsubsection{\label{sec:results_num_igw}Internal gravity waves}
\begin{figure}[h]
    \centering
    \includegraphics[scale = 0.55]{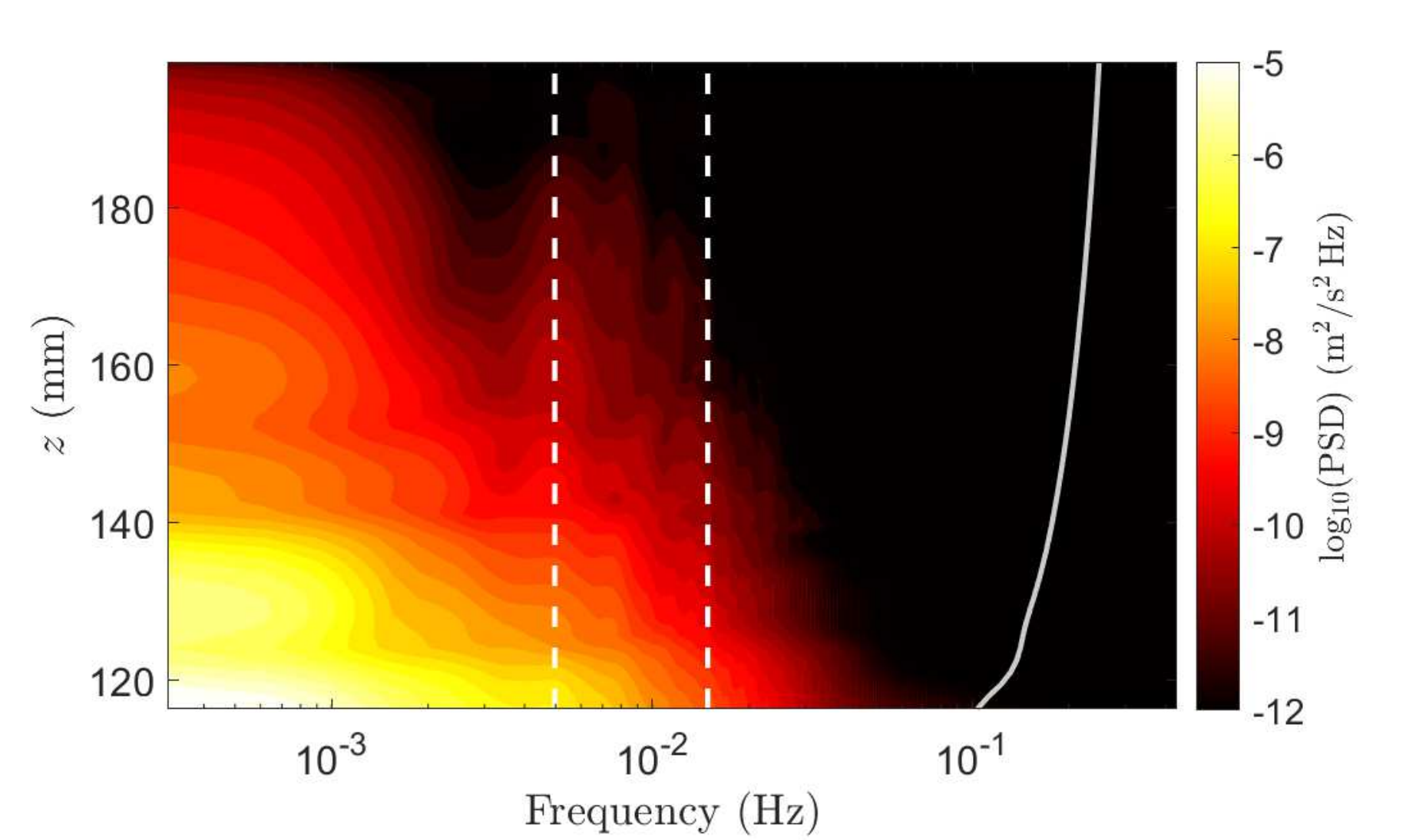}
    \hspace{.1cm}
    \includegraphics[scale = .55]{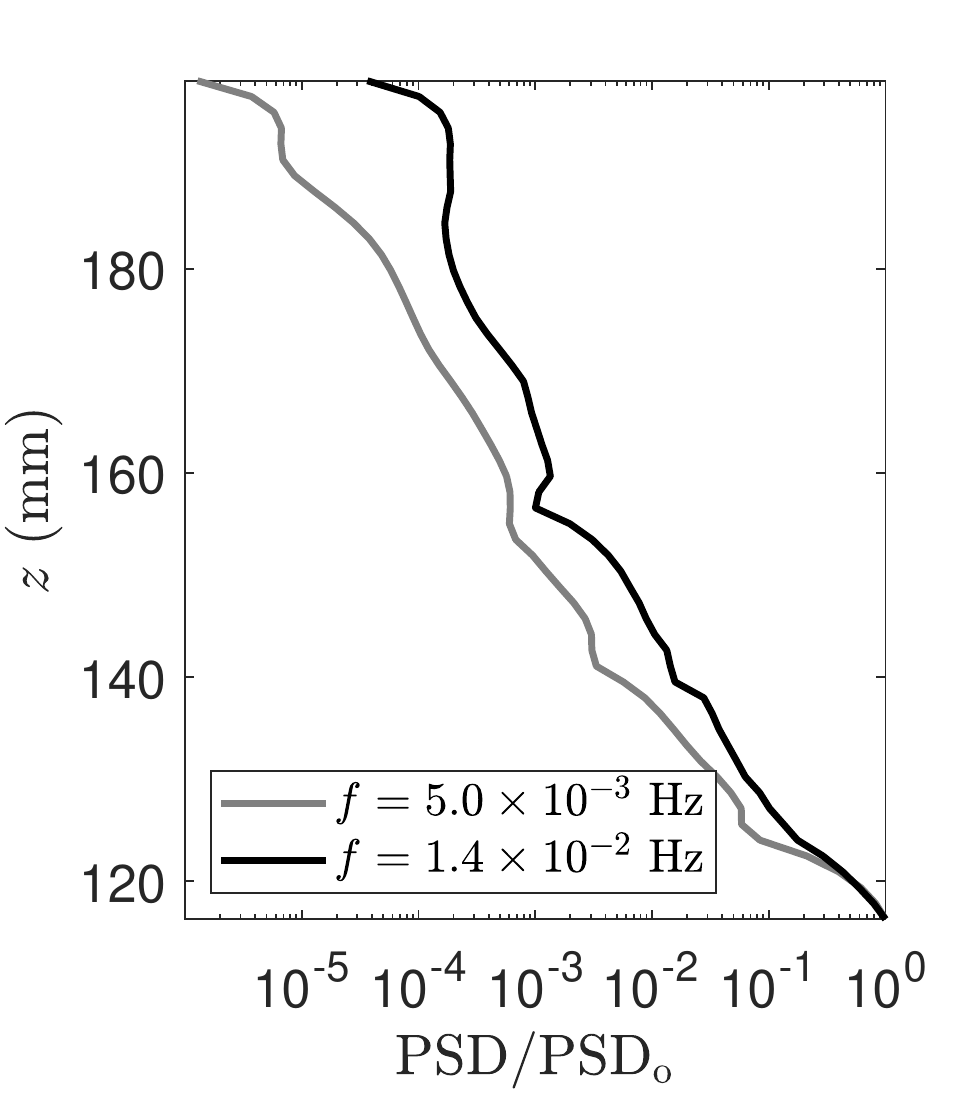}
    \caption{(Left) Power spectral density of the absolute velocity $\sqrt{u^2+w^2}$ in the buffer and stratified layers. The grey curve shows the buoyancy frequency profile computed from the spatial and temporal average of the temperature field. (Right) Two selected profiles (taken at frequencies shown by dashed lines on the left graph) of the re-scaled PSDs by the PSD at the top of the convective layer, \textit{i.e.} $z=118$~mm.}
    \label{fig:num_spectro}
\end{figure}

PSDs are computed in the stratified and buffer layers and are plotted in figure \ref{fig:num_spectro}. As for the experiment (figure \ref{fig:spectro}), numerical results show oscillatory motions at different frequencies attenuated with height. Experimental results (figure \ref{fig:spectro}) and numerical results (figure \ref{fig:num_spectro}) show strongly similar dynamics: most of the energy is present at low frequencies ($f<3 \times 10^{-3}$~Hz). The motion with frequencies ranging from $3 \times 10^{-3}$~Hz to $N$ are less intense, and almost no energy is seen at frequencies $f>N$. 

Right panel of figure \ref{fig:num_spectro} shows two selected vertical profiles (shown by the white dashed line on the left panel figure) of the the PSD re-scaled by the PSD at $z=118$~mm. The energy for the higher frequency ($f=1.4\times10^{-2}$~Hz) decreases slower than the energy for the lower frequency ($f=5.0\times10^{-3}$~Hz). This is, as experimental results, in agreement with the dispersion relation of IGWs.
The overall behaviour of waves spectra is similar in experiment and numerical simulation, with an attenuation length independent of the frequency in the low-frequency signal thus confirming a viscous coupling origin of the large-scale flow, and increasing when the frequency goes towards $N$ in the wave domain.

\subsubsection{Large-scale flow within the stratified layer}
\begin{figure*}[h]
    \centering
    \includegraphics[scale = 0.6]{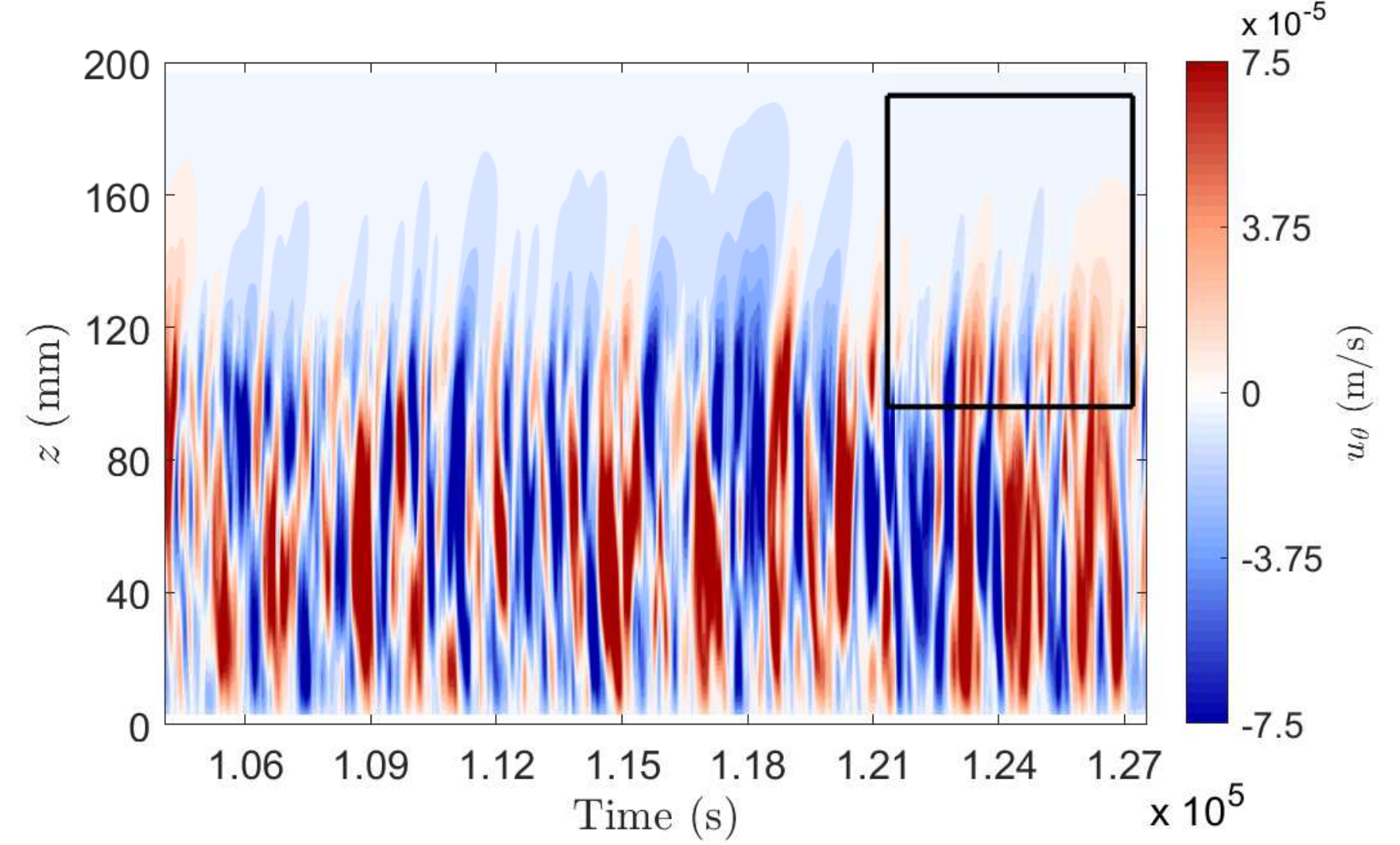}
    \hspace{.1cm}
      \includegraphics[scale = 0.6]{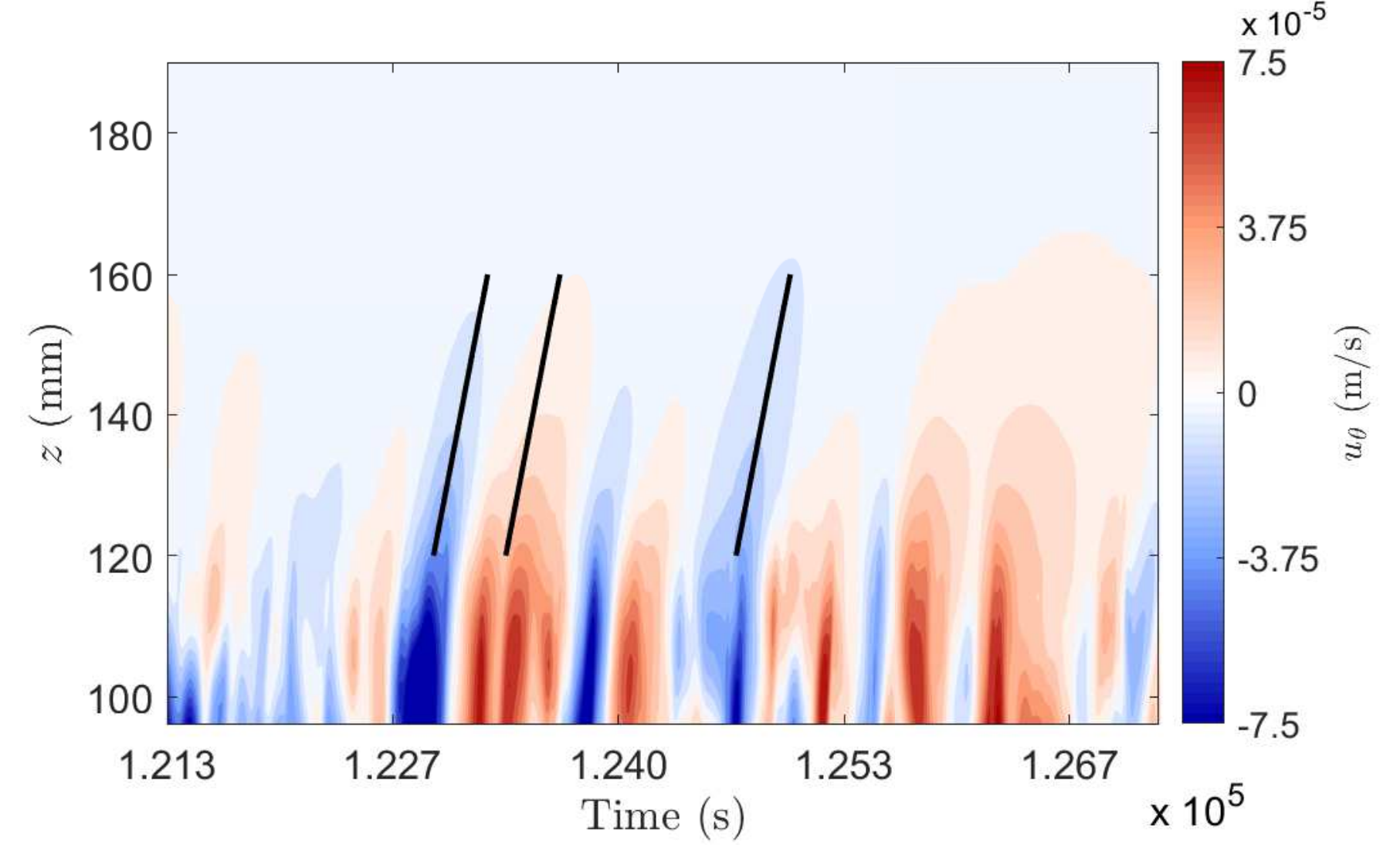}
    \caption{Spatio-temporal diagrams of the azimuthal averaging of the azimuthal velocity inside a virtual cylinder of radius $r = 140~$mm. The bottom figure is a zoom on the stratified zone, delimited in the top figure by the black square. The slope of the black lines show the theoretical viscous diffusive time.}
    \label{fig:num_vthetacyl}
\end{figure*}
Similarly to what has been done for the experimental data, figure \ref{fig:num_vthetacyl} shows the mean azimuthal velocity over the whole height of a virtual cylinder of radius $r = 140$ mm. We observe reversals within the convective layer ($z<120$ mm), which are not systematically correlated with the signal in the stratified layer. The mean velocity in the stratified layer also exhibits reversals. They are characterised by an upward phase propagation from the buffer zone at $z=120$ mm, as shown in the zoom (bottom panel of figure \ref{fig:num_vthetacyl}). The phase velocity seen in figure \ref{fig:num_vthetacyl} is in good agreement with the theoretical time for viscous propagation $t = \frac{z^2}{4\,\nu\, \mathrm{erf}^{-2}\left(\left(\frac{u}{u_b}-1\right)\right)}$. This corroborates the fact that the reversals observed within the stratified layer are viscously driven from the dynamics occurring in the buffer layer, as it has been seen for the experiment. Reversals time ranges from $300$~s to $1500$~s. Those reversals times are similar to the experimental ones, though slightly shorter (numerical reversals are $\sim20$\% faster than experimental reversals). 

\clearpage
\section{Conclusion}\label{sec:discussion}
The 4$^{\circ}$C convection experiment, originally performed by Townsend \cite{townsend_natural_1964}, has been re-investigated using long-term PIV measurements in a vertical cross-section, and in several horizontal cross-sections within the stratified layer. This last type of measurements has allowed to investigate for the first time the long-term horizontal mean flow in the stratified layer. Experiments have been complemented by direct numerical simulations. The first result of this paper is the confirmation, in 3D and with ideal boundary conditions, of the presence of a buffer layer, including an overshooting region as first observed by Perrard \cite{perrard_experimental_2013}, and a shear region. We have argued that the buffer layer is driven by thermal coupling with the convection, due to the non-linear equation of state of water, and that this mechanism is a priori related to a Prandtl number larger than one.  The second result is that the buffer layer viscously drives slow reversals of the horizontal large-scale flow within the stratified layer.

Additionally, IGWs at different frequencies propagate in the stratified layer. They likely interact with the horizontal large-scale flow, and probably also produce a reversing flow, which superimposes to the viscously driven one. From Couston et al. \cite{couston_order_2018}, we know that the Prandtl number has a strong influence on this QBO-like mechanism: the lower the Prandtl number, the stronger the amplitude of the QBO. In water, $Pr \sim 7$, and the expected amplitude of the large-scale QBO flow is weak, hence dominated by the viscous driving. Further experimental studies at lower Prandtl number should allow deciphering the two contributions. One could for instance suggest using gas as a working fluid; however, the absence of density reversal around a given temperature will necessitate to consider either transient experiments like \cite{deardorff_laboratory_1969, michaelian_coupling_2002}, or two-gas experiments which might then be prone to double diffusive instabilities. Experimentally, the question also remains to understand why the only successful QBO experiment has been performed in salty water, hence with a Schmidt number (equivalent to Pr) of 700.

\begin{figure}[h]
    \centering
    \includegraphics[scale=.6]{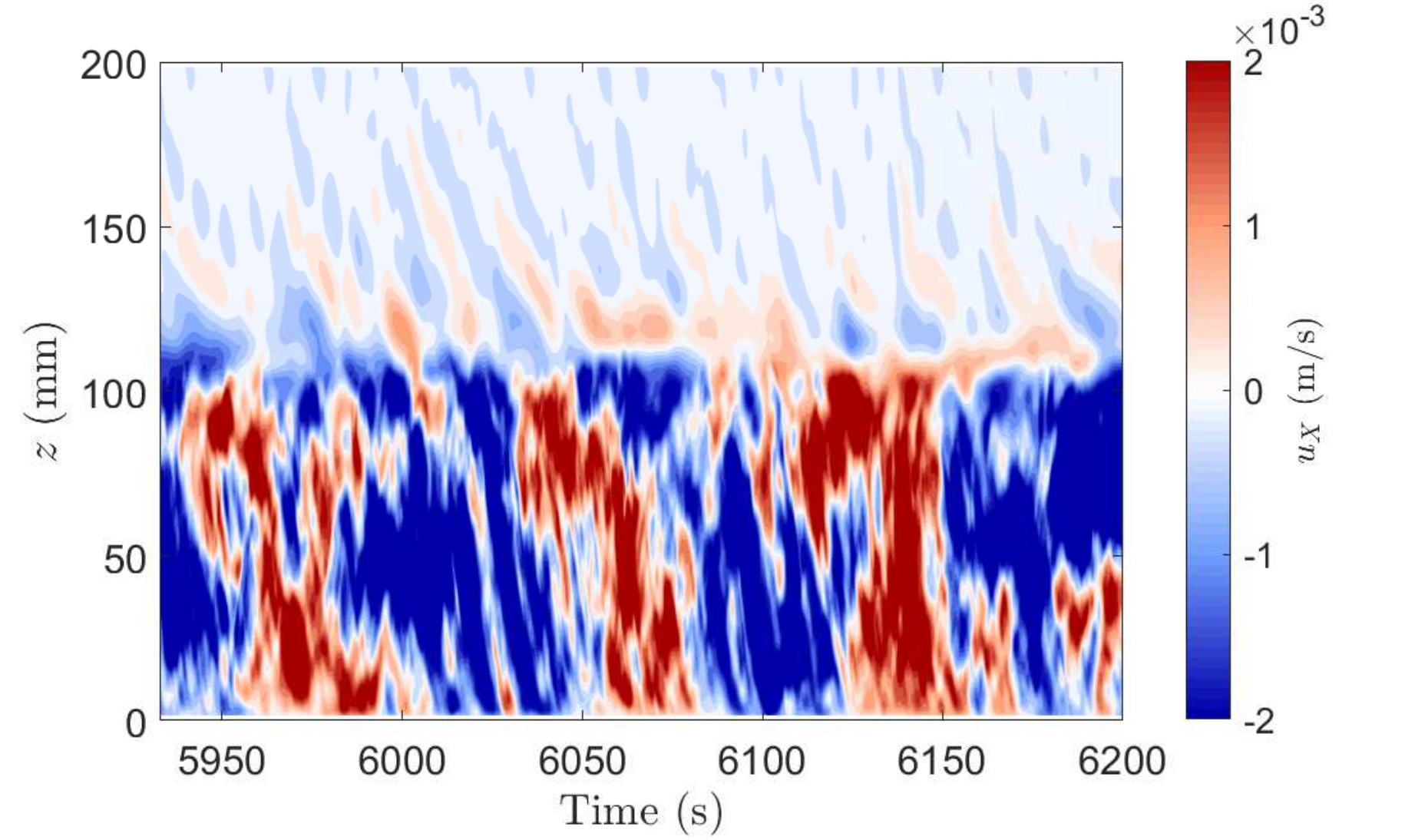}
    \caption{Horizontal average of the horizontal velocity $u$ over a vertical cross section in the middle of the numerical domain for $Pr = 0.1$.}
    \label{fig:num_pr01}
\end{figure}

In the meantime, it is straightforward to change the Prandtl number in the numerical simulation of our set-up. We have thus run a second simulation with the same Rayleigh number $Ra = 10^7$ and top temperature $\theta_{top} = 11$ but with $Pr = 0.1$. In this simulation, as shown in figure \ref{fig:num_pr01}, no buffer layer is observed, but strong signatures of a QBO like mechanism are visible, marked by downward phase propagation of the reversals of the large-scale flow. This configuration thus deserves a more systematic study in the future.

\section*{Acknowledgements} 
The authors acknowledge funding by the European Research Council under the European Union's Horizon 2020 research and innovation program through Grant No. 681835-FLUDYCO-ERC-2015-CoG. This work was granted access to the HPC resources of Aix-Marseille Universit\'e financed by the project Equip@Meso (ANR-10-EQPX-29-01) of the program “Investissements d'Avenir” supervised by the Agence Nationale de la Recherche. Computations were also conducted with the support of the HPC resources of GENCI-IDRIS (Grant No.A0060407543).

\end{document}